\documentclass[prd,superscriptaddress,amsfonts,amssymb,amsmath,showpacs,twocolumn]{revtex4-2}
\usepackage{bm}
\usepackage{amsfonts}
\usepackage{braket}
\usepackage{latexsym}
\usepackage[latin1]{inputenc}
\usepackage{graphicx}
\usepackage{amsmath}
\usepackage{mathtools}
\usepackage{palatino}
\usepackage{mathpazo}
\usepackage{textcomp}
\usepackage{float}
\usepackage{booktabs}
\usepackage{dcolumn}
\usepackage{hyperref}
\usepackage{subfigure}
\usepackage[english]{babel}
\usepackage[autostyle]{csquotes}
\hypersetup{colorlinks,citecolor=red}
\usepackage{amsmath}
\usepackage{xcolor}
\usepackage{orcidlink}
\usepackage{commath}

\def\jnl@style{\it}
\def\aaref@jnl#1{{\jnl@style#1}}

\def\aaref@jnl#1{{\jnl@style#1}}

\def\aj{\aaref@jnl{AJ}}                   
\def\apj{\aaref@jnl{ApJ}}                 
\def\apjl{\aaref@jnl{ApJ}}                
\def\apjs{\aaref@jnl{ApJS}}               
\def\apss{\aaref@jnl{Ap\&SS}}             
\def\aap{\aaref@jnl{A\&A}}                
\def\aapr{\aaref@jnl{A\&A~Rev.}}          
\def\aaps{\aaref@jnl{A\&AS}}              
\def\mnras{\aaref@jnl{Mon.~Not.~Roy.~Astron.~Soc.}}             
\def\prd{\aaref@jnl{Phys.~Rev.~D}}        
\def\prc{\aaref@jnl{Phys.~Rev.~C}}  
\def\prl{\aaref@jnl{Phys.~Rev.~Lett.}}    
\def\qjras{\aaref@jnl{QJRAS}}             
\def\skytel{\aaref@jnl{S\&T}}             
\def\ssr{\aaref@jnl{Space~Sci.~Rev.}}     
\def\zap{\aaref@jnl{ZAp}}                 
\def\nat{\aaref@jnl{Nature}}              
\def\aplett{\aaref@jnl{Astrophys.~Lett.}} 
\def\apspr{\aaref@jnl{Astrophys.~Space~Phys.~Res.}} 
\def\physrep{\aaref@jnl{Phys.~Rep.}}      
\def\physscr{\aaref@jnl{Phys.~Scr}}       
\def\commat{\aaref@jnl{Comm.~Math.~Phys.}}              
\def\science{\aaref@jnl{Science}}               
\def\cqg{\aaref@jnl{Classical Quant.~Grav.}}            
\def\jpcs{\aaref@jnl{JPCS}}                                     
\def\ijmpd{\aaref@jnl{Int.~J.~Mod.~Phys.~D}}                    
\def\grg{\aaref@jnl{Gen.~Relat.~Gravit.}}               
\def\rpp{\aaref@jnl{Rep.~Prog.~Phys.}}          
\def\npa{\aaref@jnl{Nucl.~Phys.~A}}        
\def\lrr{\aaref@jnl{Living Rev.~Rel.}}                   
\def\jcap{\aaref@jnl{J.~Cosmology Astropart.~Phys.}}    
\def\rmp{\aaref@jnl{Rev.~Mod.~Phys.}}   
\def\epjc{\aaref@jnl{Eur.~Phys.~J.~C}}

\allowdisplaybreaks[1]

\begin{document}

\color{black}

\title{Charged gravastar model in noncommutative geometry under $f(\mathbb{T})$ gravity}

\author{Debasmita Mohanty\orcidlink{0009-0006-8118-5327}}
\email{newdebasmita@gmail.com}
\affiliation{Department of Mathematics, Birla Institute of Technology and
Science-Pilani,\\ Hyderabad Campus, Hyderabad-500078, India.}
\author{Sayantan Ghosh\orcidlink{0000-0002-3875-0849}}
\email{sayantanghosh.000@gmail.com}
\affiliation{Department of Mathematics, Birla Institute of Technology and Science-Pilani,\\ Hyderabad Campus, Hyderabad-500078, India.}
\author{P.K. Sahoo\orcidlink{0000-0003-2130-8832}}
\email{pksahoo@hyderabad.bits-pilani.ac.in}
\affiliation{Department of Mathematics, Birla Institute of Technology and
Science-Pilani,\\ Hyderabad Campus, Hyderabad-500078, India.}

\date{\today}

\begin{abstract}
In this article, we study the properties of charged gravastars in torsion-based $f(\mathbb{T})$ gravity in the presence of noncommutative geometry. We have taken the interior from noncommutative motivated space-time, noting that why, from a physical point of view, such a choice is justified, then we have taken the thin shell as stiff matter and taken three different exterior metrics (Reissner-Nordstrom (R-N), Bardeen and Ayon-Beato-Garcia (ABG) metric) to construct the gravastar model. We have studied the physical properties like proper length,  entropy, energy, and EoS for these models, and we have also used Israel junction conditions to study the effective pressure, energy density, and potential of the thin shell. Finally, we comment on the stability of such a thin shell and the deflection angle caused by such a thin shell, which could, in principle, be tested by future radio telescopes like the Event Horizon Telescope (EHT).

\end{abstract}
\maketitle
\textbf{Keywords:}  Gravastar, Noncommutative geometry, Stability, Stiff matter, Deflection angle, ABG metric, Bardeen metric, Reissner-Nordstrom metric.

\section{Introduction} \label{sec:I}
Gravitational Vacuum Condensate Stars, also known as gravastars in literature, are extremely lucrative alternatives for black holes. It is well known that the end state of a sufficiently heavy ($>3M_{\odot}$) star after the supernova is usually black hole. It was first shown by Oppenheimer and Snyder \cite{Oppenheimer/1939}  that a spherically symmetric pressure-less dust ball under its own gravity can form a black hole with singularity. It was later extended by Roger Penrose \cite{Penrose/1965}, who first showed that even in more general circumstances also, the possible fate of a supermassive object is black holes with singularity. The event horizon of the supermassive black hole (M87) has already been observed by the Event Horizon Telescope (ETH) \cite{Akiyama/2019}. However, one should note that the radio observations can only resolve the event horizon based on the outside data. As it is known from Birkoff's theorem, the exterior of a static spherically symmetric object is always given by the Schwarzschild solutions. So even though black holes are one of the most fascinating objects in general relativity, there are several issues for which people have looked into the alternatives of black holes; first and foremost, black holes have a non-removable singularity at $r=0$ which is a curvature singularity, that means it can be resolved by a suitable coordinate transformation. Also, as the observations can mostly probe into the exterior of the black hole, it is reasonable to expect that there could be other alternatives to the black hole paradigm. Several papers have addressed such issues in the literature. For example, alternates of black holes have been proposed in the form of the Proca star \cite{Landea/2016}, Boson stars \cite{Alcubierre/2023, Kumar/2016}, and gravastars \cite{P/2023, Mazur/2004} just to name a few.\\
Gravastar as an alternative to black holes was first proposed by Mazur and Mottola \cite{P/2023, Mazur/2004}. The key observation of the gravastar is that during the gravitational collapse, the quantum effects of the collapsing objects would be non-negligible, so the dust or other bosons might condensate to give a Bose-Einstein condensation, which could act as dark energy-like repulsive force ($\omega=-1$). Also, in surface, one assumes a thin shell made of stiff matter ($\omega=1$) to make sure that the general entropy formula for an object made of EoS ($k$) $S=\frac{4k+4}{7k+1}S_{BH}$, we note that the $S=S_{BH}$ iff $k=1$. So, one takes stiff matter ($\omega=1$) in the shell to make sure that the entropy of the above object coincides with the black hole.\\ 
The motivation for noncommutative geometry comes from the fact string theory suggests a lower bound to distance measurement, and at small length scales, coordinates behave as noncommutative operators on a D-Brane discretizing space-time. This approach discretizes space-time with a commutator of
the form $[x^\mu,x^\nu]= i \theta^{\mu\nu}$, where $\theta^{\mu\nu}$ is an anti symmetric tensor. An interesting physical
implication of this is that at small length scales, particles become smeared objects modelled
with a Gaussian or Lorentzian distribution. The geometric contribution there can be modelled as a self-gravitating source of an-isotropic fluid with an energy density profile determined
uniquely by $\theta^{\mu\nu}$ and the effective mass. Noncommutative geometry as a gravitational
source is an intrinsic property of space-time and does not depend on curvature. To this
end, wormhole solutions with viable physical properties in noncommutative backgrounds
have been discussed previously.\\
Where $\theta^{\mu\nu}$ represented as a second-order
anti-symmetric matrix of dimension $\text{length}^2$. It is analogous to the democratization of phase space by the Planck
constant. The standard concept of mass density
in the form of the Dirac delta function does not hold
in noncommutative space. Thus, instead of using the Dirac delta function, Gaussian and Lorentzian
distributions of negligible length $\sqrt{\theta}$ can 
illustrate this distribution impact. The static, spherically symmetric smeared, and particle-like gravitational
source handles the geometry of Gaussian distribution
with a noncommutative nature with the maximum mass
M retains. Meanwhile, the density capacity of the particle-like mass M can be assumed for Lorentzian distribution. However, in our case, we have not. 
In this case, the entire mass M can be regarded as a
form of the diffused unified object, and $\theta$ is the noncommutative parameter. Distribution has been used to specify
the physical substances of short-separated divergences
of noncommutative coordinates in probing black holes and other astrophysical objects like gravastars.\\
This effective approach may be considered as an improvement to semiclassical gravity and understanding the noncommutative effects. Motivated by this idea, models for
 noncommutative geometry inspired Schwarzschild-Tangherilini black holes were obtained by
Nicolini \cite{Nicolini/2006}, Smailagic \cite{Smailagic1/2003}, and Spallucci \cite{Ansoldi/2007} extended to the Reissner-Nordstrom in
four dimensions, generalized to higher dimensional space-time by Rizzo \cite{Rizzo/2006}, to charge
in higher dimensions and then to the BTZ black holes. S. G. Ghosh \cite{Ghosh/2018} has also studied noncommutative black holes in the context of Gauss-Bonnet gravity. Further, in recent years, we have witnessed significant interest in noncommutative models, mainly due to their relevance to quantum gravity. We would also like to note that the thermodynamic properties of a noncommutative 
black hole have been studied by  Banerjee et al.\cite{Banerjee/2008}. This shows that one can indeed recover the classical Bekenstein-Hawking entropy formula for the black hole in noncommutative space-time as well. A review of the noncommutative-inspired model can be found in this paper. Thus, the main effect of noncommutative is proposed to be the smearing out
of conventional mass distributions. Hence, we will take, instead of the point mass, M,
described by a $\delta$-function distribution, a static, spherically symmetric, Gaussian-smeared
matter source, in D- dimensions are given as \cite{Spallucci}:
\begin{equation}\label{eq:1}
    \rho_{\theta}(r)=\frac{\mu}{4 \pi\theta^{\frac{D-1}{2}}}e^-{\frac{r^2}{4\theta}}
\end{equation}
The particle mass $\mu$ diffused throughout a region of linear size $\theta$. Here, $\theta$ is the noncommutative parameter, which is considered to be a Planck limit. Thus, the above equation plays the role of a matter source, and the mass is smeared around the region $\sqrt{\theta}$ instead of locating at a point. In our model, we will use this in the inner gravastar to find the feasibility of the solutions.\\
The canonical idea of noncommutative geometry (from the mathematical point of view) was first proposed in \cite{Doplicher}, where there was a physically motivated argument on why, in the very high energy scales, noncommutative should be included in the calculation. Soon, it was followed by the \cite{Spallucci} path integral approach, and it has been shown that the $\theta$ indeed satisfies the Gaussian distribution.  \\
Here, we will briefly discuss how the Gaussian distribution arises due to noncommutative geometry quite naturally. \\
As it is known from the propagating kernel of the path-integral formulation, it needs to be evaluated in the small time slices like $\braket{x_{i+1},\epsilon|x_i,0}$. Putting the momentum completeness inside, one would have to evaluate the following terms in the presence of noncommutative geometry $\braket{p|x}_{\theta}$.\\
Given the set of Cartesian coordinates $x_i,x_j$ (in general, this can be extended for any generalised coordinates) we define the ``canonical noncommutativity" as $[x_i,x_j]=i\theta$ then one can define a set of new variables (taking motivation from the raising and lowering operator of the SHM) as $A=x_i+ix_j, A^\dag = x_i-ix_j $ this would give $[A, A^\dag]=2 \theta$. We also note the following identity $e^{ip_+A^\dag+ip_-A}=e^{ip_+A^\dag+}e^{ip_-A}e^{-\theta \frac{p^2}{4}}$. where $2P_+=p_i+ip_j ,\,\ 2P_-=p_i-ip_j$. \\
Finally, one can use this to find the propagation kernel as follows:
\begin{multline*}
K_\theta(x-y:T)=N \int [\mathbf{D}x][\mathbf{D}p] Exp (i\int_x^y \Vec{p}.d\Vec{x} \\-
\int_0^Td\tau (H(\Vec{p},\Vec{x})+\frac{\theta}{2T}\Vec{p}^2) )
\end{multline*}
So, Green's function in the momentum space can be written as 
$$G_\theta(p^2;m^2)=\left(\frac{1}{2\pi}\right)^3 \frac{exp(\theta p^2/ 2)}{p^2+m^2}$$
This gives the origin of the Gaussian term in the noncommutative geometry. We would also like to note that some subtle issues are present in the derivation. Most importantly, the derivation is ``non-local" in nature. However, we note that the paper by Gangopadhyay and Scholtz \cite{Gangopadhyay/2009} clarifies these issues by introducing an additional auxiliary field. However, the Gaussian nature is exactly the same as the previous derivation.

We would first like to note that the study of noncommutative geometry in the astrophysical context was first studied by Nicolini \cite{Nicolini} followed by \cite{Nicolini2}. In the review article by 
 Nicolini \cite{Nicolini}, there is a detailed discussion about the motivation for choosing noncommutativity in the black hole context. Roughly, the idea goes as follows: it is well known that beyond plank length, our idea of space-time needs to be modified to consider the non-renormalizable effects of gravity. As discussed earlier, one lucrative effect of such a paradigm is the so-called noncommutative space-time hypothesis. The idea is that such an effect can indeed be studied in the context of black holes. In the above article, the author has explicitly shown that black holes in such noncommutative spacetime geometry not only reproduce the original black hole entropy calculated by Hawking using semi-classical treatment but also provide logarithmic corrections to the entropy. This is a strong indication that if compact objects like black holes or some proposed alternative of black holes (gravastar in our context) are presented as the classical solutions to Einstein's relativity, then it is reasonable that one extend it to a noncommutative space to take some of the Trans-Planckian effects. Also, the study has been extended in the wormhole context by Rahaman \cite{Islam, Banerjee} and Hassan \cite{Hassan1}. Later, it has been further studied in the wormhole context in \cite{Tayade/2023} .\\

The first motivation for the ABG metric comes from the paper by Ayon-Beato and Garcia \cite{Eloy/1999}, where they have pondered over the fact that if one tries to find the solution of a black hole under non-linear electrodynamics. It has been shown that only for a very particular Hamiltonian (as it is very difficult to solve for the general form of the Hamiltonian). The solution is also regular and, in the limiting case, leads to R-N solutions. The study was further extended in the case of more regular black holes in \cite{Ayon/1998} and also included magnetic monopoles in \cite{Bronnikov/2001}. We also note that the properties of the interior and the horizons of such black holes have been studied in detail in \cite{Matyjasek/2013}. We have determined the gravastar-related deflection angle, which will be useful for the next phenomenological research. We also note that the reason we take the charged gravastars is because it has a close analogue to the charged hair of black holes. Also, it has been shown that Nicolini \cite{Nicolini/2018} that the charged black holes are especially interesting in the context of noncommutative geometry as they provide a very good playground for Schwinger effects by which black holes (in this case, gravastars discharges). It has also been shown that \cite{Ovgun/2020}  in Rastall gravity, the shadow of a black hole gets a correction due to noncommutative. Also, some further properties of the black holes in the context of noncommutative geometry have been studied in \cite{Araujo/2024}. \\

 The Gravastar model has been thoroughly investigated within the context of various alternative gravitational theories. In 2020, Das and his colleagues \cite{Das/2020} investigated the formation of gravastars within the context of the $f(\mathbb{T})$ gravity theory. Pradhan et al. \cite{Pradhan/2023} studied the thin-shell gravastar model in $f(Q, T)$ gravity. Mohanty et al., \cite{Mohanty} also investigate the study of the charged gravastar model in $f(Q)$ gravity.
It should be noted that although several works in past have been explored for the charged gravastars in various gravity \cite{Ovgun/2017, Lobo/2013, Debnath/2019}.\\

In a gravastar model (Gravitational Vacuum Star), the three distinct regions are characterized by different equations of state (EoS) and physical properties. Each layer is assumed to have $r_1< r_2$ and inner and outer radii of $r_1$ and $r_2$, respectively.:
\begin{itemize}
    \item Interior region: $ 0 \leq r \leq r_1$, $\omega=-1$, i:e:, EoS is $p=-\rho$ ,
    \item Shell: $r_1\leq r \leq r_2$, $\omega=1$,  i:e:, EoS is $p=\rho$ 
    \item Exterior region:  $r_2 \leq r $, EoS $p=0$.
\end{itemize}
The rest of this paper is arranged as follows. In section \ref{sec:II}, we set up the basic of the $f(\mathbb{T})$ gravity. In section \ref{sec:III}, we derive the field equation of $f(\mathbb{T})$ gravity. In section \ref{sec:IV}, we discuss the noncommutative energy density and interior region of gravastar. In section \ref{sec:V}, we present the thin shell. In section \ref{sec:VI}, we discuss the external region by taking different regular black holes. In section \ref{sec:VII}, we derive the junction condition. In section \ref{sec:VIII}, we investigate the physical behaviour of the model. In section \ref{sec:IX}, we derive the boundary condition. In section \ref{sec:X}, we analyse the deflection angle. In section \ref{sec:XI}, we explore the stability of the model. In section \ref{sec:XII}, we give some concluding remarks.

\section{Basic mathematical formalism of the $f(\mathbb{T})$ theory} \label{sec:II}

One may define any metric as $g_{\mu \nu}=\eta_{ij} \, e^i\,_{\mu} \, e^j\,_{\nu}$ with $\eta_{ij}=diag (1,-1,-1,-1)$ and $e_i^{\mu} e^i_{\nu}=\delta^{\mu}_{\nu}$, $e=\sqrt{-g}=det [e^i\,_{\mu}]$. Here $e^i\,_{\mu}$ denotes the tetrad fields.\\

Where the $\mathbb{T}^{\sigma} \,_{\mu \nu}$ and $ K^{\mu \nu}\,_{\sigma}$ denotes the torsion and contorsion tensor respectively, which is given by 

\begin{equation} \label{eq:2}
     \mathbb{T}^{\sigma} \,_{\mu \nu}= \Tilde{\Gamma^{\sigma}}\,_{\mu \nu}-\Tilde{\Gamma^{\sigma}}\,_{\nu \mu}=e_i^{\sigma}\left( \partial_{\mu} \, e^i \,_{\nu}-\partial_{\nu} \, e^i\,_{\mu}  \right), 
 \end{equation}
 \begin{equation}\label{eq:3}
    K^{\mu \nu}\,_{\sigma}=-\frac{1}{2} \left(\mathbb{T}^{\mu \nu}\,_{\sigma}-\mathbb{T}^{\nu \mu}\,_{\sigma}-\mathbb{T}_{\sigma}\,^{\mu \nu}  \right).
\end{equation}

The components of tensor $S_{\sigma}\,^{\mu \nu}$ can be defined as
\begin{equation}\label{eq:4}
 S_{\sigma}\,^{\mu \nu}=\frac{1}{2} \left( K^{\mu \nu}\,_{\sigma}+ \delta^{\mu}_{\sigma} \,\mathbb{T}^{\beta \nu} \,_{\beta} - \delta^{\nu}_{\sigma} \, \mathbb{T}^{\beta \mu}\,_{\beta}
 \right).
\end{equation}
The torsion scalar is defined as:

\begin{equation}\label{eq:5}
    \mathbb{T}=S_{\sigma}\,^{\mu \nu} \,\mathbb{T}^{\sigma}\,_{\mu \nu} .
\end{equation}

The action of $f(\mathbb{T})$ theory, as stated in references \cite{Bengochea/2009, Baojiu/2011}, is defined under the assumption of geometrized units where $G = c = 1$.

\begin{equation} \label{eq:6}
    S[e^i_{\mu},matter]= \int d^4 x \,\,e \left[\frac{1}{16 \pi} f(\mathbb{T})+\mathcal{L}_{matter}\right],
\end{equation}

where $f(\mathbb{T})$ is an arbitrary analytic function of the torsion scalar $\mathbb{T}$. Also, it reduces to GR when $f(\mathbb{T})=\mathbb{T}$.





The field equations of $f(\mathbb{T})$ gravity are produced by varying the action \eqref{eq:6} with respect to the tetrad \cite{Bengochea/2009, Baojiu/2011, Bohmer/2011} as

\begin{multline} \label{eq:7}
    S_i\,^{\mu \nu} \, f_{\mathbb{TT}} \,\partial_{\mu} \mathbb{T} + e^{-1} \partial_{\mu} (e S_i\,^{\mu \nu}) f_{\mathbb{T}} -\mathbb{T}^{\sigma}\,_{\mu i} \, S_{\sigma}\,^{\nu \mu} \, f_{\mathbb{T}} +\frac{1}{4} e_i^{\nu} f \\
    = 4 \pi T_i^{\nu},
\end{multline}

where 
\begin{align} \label{eq:8}
    S_i\,^{\mu \nu}= e_i\,^{\sigma} S_{\sigma}\,^{\mu \nu}, \,\,\,\,\, f_{\mathbb{T}}=\frac{\partial f}{\partial \mathbb{T}}, \,\,\,\, f_{\mathbb{TT}}=\frac{\partial^2 f}{\partial \mathbb{T}^2}.
\end{align}

And $T_i^{\nu}$ denotes the energy stress tensor of the perfect fluid, which is given by

\begin{equation} \label{eq:9}
    T_{\mu \nu}=(\rho + p_t) u_{\mu}\, u_{\nu} - p_t g_{\mu \nu}+(p_r-p_t)v_{\mu} v_{\nu}.
\end{equation}

In the expression, where $u_\mu u_\nu =-v_{\mu} v_{\nu}=1$, $u_\mu$,  and $v_{\mu}$ represent the four-velocity vectors, while $\rho$ and $p$ denote the matter-energy density and the fluid pressure of the system, respectively.

\section{The field equations in $f(\mathbb{T})$ gravity} \label{sec:III}
Let us consider a spherically symmetric line element
with metric signature $(+,-,-,-)$:

\begin{equation} \label{eq:10}
ds^2= e^{\nu(r)} dt^2 - e^{\lambda(r)} dr^2 -r^2(d \theta^2 + \sin^2 \theta \, d\phi^2),
\end{equation}
where the metric potential $e^{\nu(r)}$ and $e^{\lambda(r)}$ are the functions of radial coordinate $r$ only.

One can define the tetrad matrix $[e^i\,_{\mu}]$ as follows for the invariance of the line element under the Lorentz transformation:

\begin{equation}\label{eq:11}
    [e^i\,_{\mu}]= \begin{pmatrix}
        e^\frac{{\nu(r)}}{2} &  0  &  0  &  0 \\
        0                    &  e^\frac{\lambda(r)}{2}  &   0   &    0    \\
         0    &   0  &    r  &   0  \\
         0    &    0    &    0    &   r \sin \theta
    \end{pmatrix}
\end{equation}
and $e=det(e^i\,_{\mu})=e^\frac{\nu+\lambda}{2}r^2 \sin{\theta}$ .\\
The non-zero torsion tensor are
\begin{equation}\label{eq:12}
    \begin{gathered}
        T^0\,_{01}=\frac{\nu'}{2} e^{\nu},\, T^0\,_{10}=-\frac{\nu'}{2} e^{\nu},\, T^2\,_{12}=-r,\\
        T^2\,_{21}=r, T^3\,_{13}=-r \sin^2 \theta ,\,T^3\,_{31}=r \sin^2 \theta,\\
        T^3\,_{23}=-r^2 \sin{\theta} \cos{\theta},\, T^3\,_{32}= r^2 \sin{\theta} \cos{\theta}.
    \end{gathered}
\end{equation}

And the non-zero contorsion tensor are
\begin{equation}
\begin{gathered}\label{eq:13}
   K^{01}\,_0=-\frac{\nu'}{2} e^{\nu-\lambda},\, K^{10}\,_0=\frac{\nu'}{2} e^{\nu-\lambda},\\
   K^{12}\,_2= r e^{-\lambda},\, K^{21}\,_2=- r e^{-\lambda},\\
   K^{13}\,_3= r e^{-\lambda} \sin^2 \theta,\, K^{31}\,_3=-r e^{-\lambda} \sin^2 \theta, \\
   k^{23}\,_3= \sin{\theta} \cos{\theta}, \, k^{32}\,_3 =-\sin{\theta} \cos{\theta}.
   \end{gathered}
\end{equation}

The corresponding torsion scalar is given by

\begin{equation}\label{eq:14}
    \mathbb{T}(r)=\frac{2 e^{-\lambda}}{r} \left(\nu' + \frac{1}{r}\right).
\end{equation}
The prime $(')$ in the formula above stands for the derivative with respect to the radial coordinate $r$.

The field equations can be obtained by inserting the components of $S_i\,^{\mu \nu}$ and $T^i\,_{\mu \nu}$ in Eq. \eqref{eq:7}.

\begin{equation} \label{eq:15}
    4 \pi \rho =-\frac{e^{-\lambda}}{r} \mathbb{T}' \, f_{\mathbb{TT}} - \left[ \frac{\mathbb{T}}{2}- \frac{1}{2 r^2} - \frac{e^{-\lambda}}{2r}(\nu' + \lambda')\right] f_{\mathbb{T}} + \frac{f}{4},
\end{equation}

\begin{equation} \label{eq:16}
    4 \pi p_r = \left[\frac{\mathbb{T}}{2} - \frac{1}{2 r^2}   \right] f_{\mathbb{T}} - \frac{f}{4},
\end{equation}

\begin{multline} \label{eq:17}
    4 \pi p_t = \frac{e^{-\lambda}}{2} \left(\frac{\nu'}{2}+\frac{1}{r}\right) \mathbb{T}' f_{\mathbb{TT}} + \frac{\mathbb{T}}{4} f_{\mathbb{T}} + \frac{e^{-\lambda}}{2} \left[\frac{\nu"}{2} + \right. \\ \left.
    \left(\frac{\nu'}{4}+\frac{1}{2r}\right) (\nu' - \lambda')\right]f_{\mathbb{T}} - \frac{f}{4},
\end{multline}

\begin{equation} \label{eq:18}
    \frac{e^{-\frac{\lambda}{2} \cot{\theta}}}{2r^2} \mathbb{T}' f_{\mathbb{TT}}=0.
\end{equation}

The field equation \eqref{eq:18} requires off-diagonal terms to be satisfied. Therefore, we can only consider $f(\mathbb{T})$ in the following form:

\begin{equation} \label{eq:19}
    f(\mathbb{T}) = a \mathbb{T} + b .
\end{equation}
Where $a$ and $b$ are constants which consequently implies that $f_{\mathbb{T}}=a$ and $f_{\mathbb{TT}}$=0.

Now, inserting Eq. \eqref{eq:19} in the equations \eqref{eq:15}- \eqref{eq:17} our field equations becomes 

\begin{equation} \label{eq:20}
    4 \pi \rho = \frac{a}{2r} \left[\lambda' \, e^{-\lambda} - \frac{e^{-\lambda}}{r}  + \frac{1}{r} \right] + \frac{b}{4},
\end{equation}

\begin{equation} \label{eq:21}
    4 \pi p_r = \frac{a}{2r} \left[\nu' \, e^{-\lambda}+ \frac{e^{-\lambda}}{r}  - \frac{1}{r} \right] - \frac{b}{4},
\end{equation}

\begin{equation} \label{eq:22}
    4 \pi p_t =\frac{a e^{-\lambda}}{2} \left[ \frac{\nu"}{2}+ \left(\frac{\nu'}{4}+\frac{1}{2r}\right) \left(\nu' - \lambda'\right)\right]-\frac{b}{4}.
\end{equation}

In addition, the conservation equation for spherically symmetric static space-time in $f(\mathbb{T})$ theory \cite{Bohmer/2011} is as follows:

\begin{equation}\label{eq:23}
    4 \pi p' + 2 \pi \, \nu' (\rho + p)= - \frac{\mathbb{T}'}{2 r^2} f_{\mathbb{TT}}.
\end{equation}

Again, by using Eq. \eqref{eq:19} in Eq. \eqref{eq:23} the above conservation equation becomes
\begin{equation} \label{eq:24}
    \frac{dp}{dr}+ \frac{\nu'}{2}(\rho+p)=0.
\end{equation}

\section{Interior (noncommutative geometry)} \label{sec:IV}
It should be noted that the metric for the Schwarzschild black hole with noncommutative geometry was first given by Smailagic et al. \cite{Spallucci}, and it was later extended for the charges  noncommutative black holes by Ansoldi et al. \cite{Ansoldi/2007}.\\
The metric for the charged $Q$ system with non-commutative geometry is given as:
\begin{equation} \label{eq:25}
     ds^2=e^{2\Phi(r)} dt^2 -\frac{dr^2}{1-\frac{2m(r)}{r}+\frac{Q_{\alpha}^2(r)}{r^2}}-r^2 d\theta^2 - r^2 \sin^2 \theta d \phi^2 .
\end{equation}
Here $m(r)$, $\Phi(r)$ and $Q_{\alpha}(r)$ are arbitrary functions of the radial coordinate $r$ only. $m(r)$ represents the quasi-local mass, which is also referred to as mass function and is defined by

\begin{equation}\label{eq:26}
     m(r)= \frac{2M}{\sqrt{\pi}} \gamma \left(\frac{3}{2},\frac{r^2}{4 \alpha} \right).
\end{equation}

The charge function $Q_{\alpha}(r)$ defined by
\begin{equation}\label{eq:27}
  Q_{\alpha}(r)=\frac{Q}{\sqrt{\pi}} \sqrt{\gamma^2 \left(\frac{1}{2}, \frac{r^2}{4 \alpha}\right)- \frac{r}{\sqrt{2 \alpha} }\gamma \left(\frac{1}{2}, \frac{r^2}{2 \alpha}\right)},
\end{equation}

and
 \begin{equation}\label{eq:28}
     \gamma \left(\frac{a}{b},x \right)=\int^{x}_{0} t^{\frac{a}{b}-1} \, e^{-t} dt.
 \end{equation}

By comparing the line element \eqref{eq:25} and \eqref{eq:10} through the field equations \eqref{eq:20}-\eqref{eq:22}, we obtained the field equations as

\begin{multline}\label{eq:29}
    \rho = \frac{1}{16 \pi  r^2 \left(-2 r m(r)+Q(r)^2+r^2\right)^2}\left[-4 a r^4 m'(r) \right. \\ \left. -8 a r m(r) Q(r)^2  
    +8 a r^2 m(r)^2+4 a r^3 Q(r) Q'(r)-2 a r^2 Q(r)^2 \right.\\ \left. +2 a Q(r)^4-4 b r^3 m(r) Q(r)^2-4 b r^5 m(r)+4 b r^4 m(r)^2  \right.\\ \left. +2 b r^4 Q(r)^2  +b r^2 Q(r)^4+b r^6\right],
\end{multline}

\begin{multline} \label{eq:30}
    p_r=\frac{1}{16 \pi  r^2 \left(-2 r m(r)+Q(r)^2+r^2\right)} \left[4 a r m(r)-2 a Q(r)^2 \right.\\ \left. +4 a r^3 \phi '(r)+2 b r^3 m(r)-b r^2 Q(r)^2-b r^4\right],
\end{multline}

\begin{equation} \label{eq:31}
  p_t=  \frac{\frac{a r^2 \left(\frac{\left(r \phi '(r)+1\right) \left(\frac{r \left(r m'(r)-m(r)-Q(r) Q'(r)\right)+Q(r)^2}{r \left(-2 r m(r)+Q(r)^2+r^2\right)}+\phi '(r)\right)}{r}+\phi ''(r)\right)}{2 \left(-2 r m(r)+Q(r)^2+r^2\right)}-\frac{b}{4}}{4 \pi }.
\end{equation}

Using the field equations \eqref{eq:29} and \eqref{eq:30} and the equation of state $p_{r}=-\rho $, we get the following relationship.
\begin{equation}\label{eq:32}
    \Phi '(r)=\frac{r^2 m'(r)-r m(r)-r Q(r) Q'(r)+Q(r)^2}{r \left(-2 r m(r)+Q(r)^2+r^2\right)},
\end{equation}
which provides 
\begin{equation}\label{eq:33}
\Phi (r)= -\frac{1}{2} \log \left[-2 r m(r)+Q(r)^2+r^2\right]+\log (r)+ \log (c_1),
\end{equation}
where $\log (c_1)$ is the integrating constant.\\
so we get 
   \begin{equation}\label{eq:34}
\Phi (r)= \frac{1}{2} \log \left(\frac{r}{ \left[-2 r m(r)+Q(r)^2+r^2\right]}\right) +\log (c_1).
\end{equation}
We note that as we impose the asymptotic flatness condition for the interior metric (otherwise, we can not interpret the $m(r)$ as ADM mass). So at $r \rightarrow \infty$, $\Phi \rightarrow 0$, so we get $c_1=0$. 
so the line elements become,
\begin{equation} \label{eq:35}
     ds^2=e^{2\Phi(r)} dt^2 -\frac{dr^2}{1-\frac{2m(r)}{r}+\frac{Q_{\alpha}^2(r)}{r^2}}-r^2 d\theta^2 - r^2 \sin^2 \theta d \phi^2 .
\end{equation}

\begin{figure}[h]
    \centering
    \includegraphics[scale=0.41]{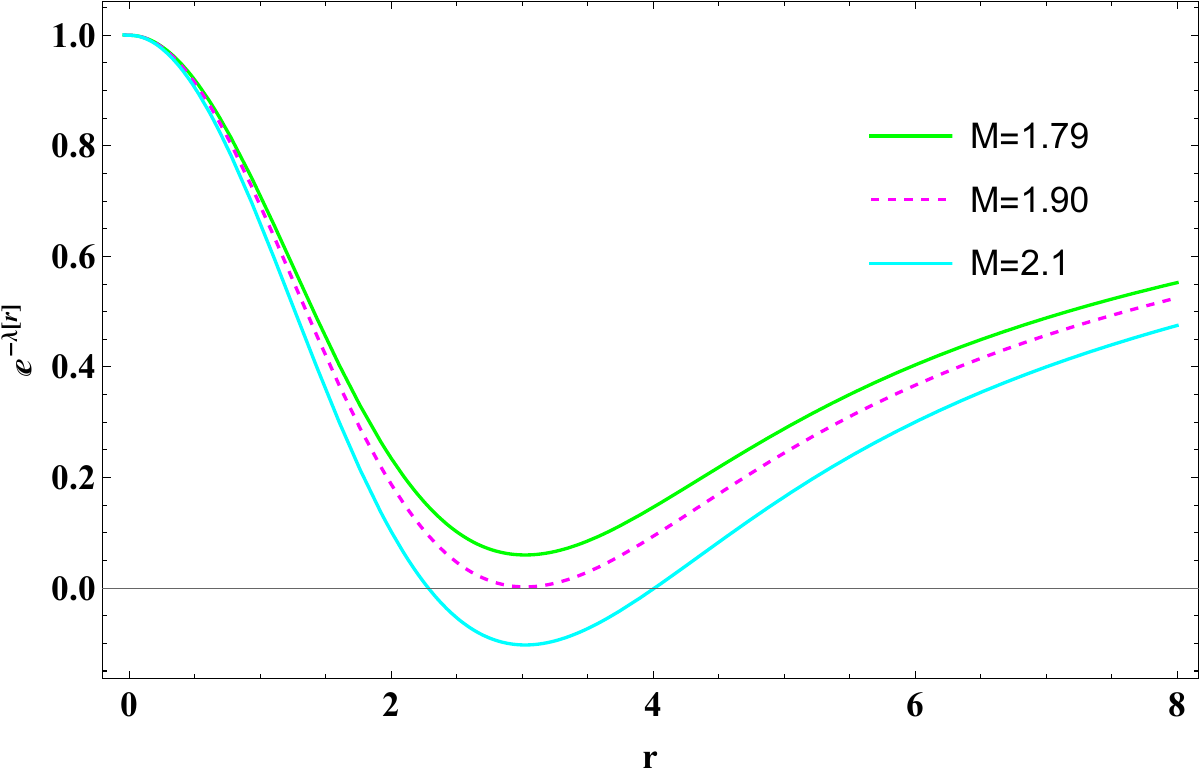}
    \caption{Plot represents the metric potential ($1-\frac{2m(r)}{r}+\frac{Q_{\alpha}^2(r)}{r^2}$) for various values of $M$ for $Q=0$.}
    \label{fig:1}
\end{figure}

\begin{figure}[h]
    \centering
    \includegraphics[scale=0.41]{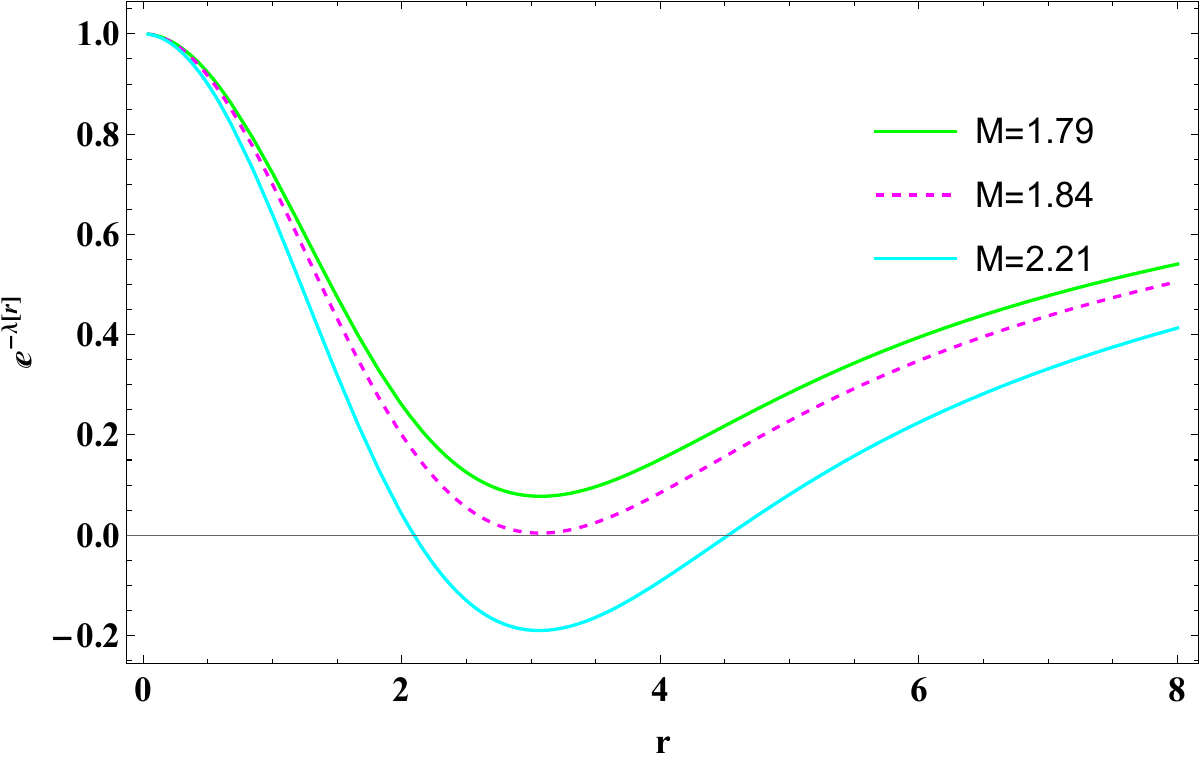}
    \caption{Plot represents the metric potential $1-\frac{2m(r)}{r}+\frac{Q_{\alpha}^2(r)}{r^2}$  for various values of $M$ for $Q=1$.}
    \label{fig:2}
\end{figure}

\begin{figure}[h]
    \centering
    \includegraphics[scale=0.41]{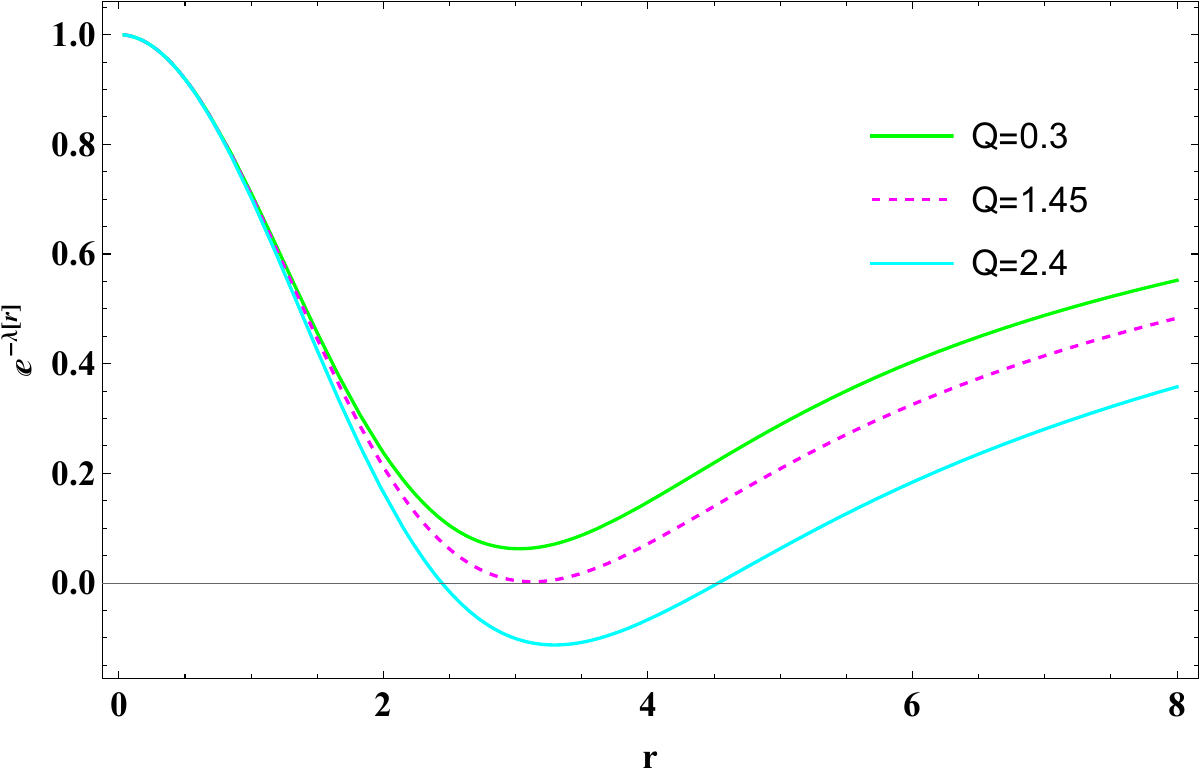}
    \caption{Plot represents the metric potential $1-\frac{2m(r)}{r}+\frac{Q_{\alpha}^2(r)}{r^2}$ for various values of $Q$ for $M=1.78$.}
    \label{fig:3}
\end{figure}

\section{shell} \label{sec:V}
The shell follows the EOS $p=\rho$ and is composed of ultra-relativistic fluid. This type of fluid, also referred to as the stiff fluid, was initially conceptualised by Zel'dovich \cite{Zeldovich/1972} in relation to the cold baryonic world. In this instance, the cause might be either the preserved number density of the gravitational quanta at zero temperature or the thermal excitations with insignificant chemical potential 
\cite{P/2023, Mazur/2004}. Many cosmological \cite{Madsen/1992} and astrophysical
\cite{Carr/1975, Braje/2002, Rahaman/2014, Mohanty/2024} phenomena have been studied using this kind of fluid. The non-vacuum region, or the shell, is where one may witness how hard it is to solve the field equations. But within the parameters of the thin shell limit, that is, $0 < e^{-\lambda}<<1$, it is possible to find an analytical solution. According to Israel \cite{Israel/1966}, we could contend that the space-time intermediate region, which is the vacuum interior and the Schwarzschild outside in this instance, must be a thin shell. Generally, any parameter that depends on $r$ is $<< 1$ as $r \to 0$ (this also applies to the thin shell region). The following equations can be obtained using this approximation: the aforementioned EOS and Eqs. \eqref{eq:20}, \eqref{eq:21}, and \eqref{eq:22}.
\begin{equation}\label{eq:36}
    4\pi \rho=\frac{a}{2r}\left(\lambda' e^{-\lambda}-\frac{e^{-\lambda}}{r}+\frac{1}{r}\right)+\frac{b}{4},
\end{equation}

\begin{equation} \label{eq:37}
    4 \pi p_r = \frac{a}{2r} \left(\frac{e^{-\lambda}}{r}-\frac{1}{r}\right)-\frac{b}{4},
\end{equation}

\begin{equation} \label{eq:38}
    4 \pi p_t=\frac{a e^{-\lambda}}{2}\left(-\frac{\nu' \lambda'}{4}-\frac{\lambda'}{2r}\right)-\frac{b}{4}.
\end{equation}
By solving \eqref{eq:36} and \eqref{eq:37} we obtain,
\begin{equation} \label{eq:39}
    e^{-\lambda (r)}=1+\frac{b \left(r^4-16 c_1\right)}{4 a r^2},
\end{equation}
where $c_1$ is an integrating constant. Using \eqref{eq:39} in the above equation, we get,
\begin{equation}\label{eq:40}
  e^{\nu(r)}=  e^{c_2} \left(1+\frac{16 c_1}{r^4}\right).
\end{equation}

\section{External region} \label{sec:VI}
As for the exterior geometry, we select two different regular BHs.
\begin{equation}\label{eq:41}
    ds^2= F(r) dt^2 - F(r)^{-1} dr^2 - r^2 d{\theta}^2 - r^2 \sin^2 \theta d{\phi}^2.
\end{equation}

\begin{itemize}
\item $F(r)$ represents R-N metric if,
$F(r)=1-\frac{2M}{r}+\frac{Q^2}{r^2}$.

\item $F(r)$ represents Ayon-Beato-Garcia (ABG) metric if,
$F(r)=1-\frac{2M}{r}+\frac{2M}{r} \tanh{\left(\frac{Q^2}{2Mr}\right)}$.

\item $F(r)$ represents charged Bardeen black holes if,
$F(r)=1-\frac{2Mr^2}{(r^2+Q^2)^{\frac{3}{2}}}+\frac{Q^2 r^2}{(r+2+Q^2)^2}$ .

\end{itemize}

\subsection{R-N metric}
Reissner-Nordstrom metric is the solution for the exterior of charged black holes. Even though for most practical purposes, the charge of astrophysics excretes is taken to be zero, there are possibilities of Hair in the black hole solutions, which are discussed in (charge hair black hole), to have. So, with that motivation, we are taking canonical R-N solutions in the exterior.
\begin{multline} \label{eq:42}
ds^2=\left( 1-\frac{2M}{r}+\frac{Q^2}{r^2}\right) dt^2 -\frac{dr^2}{ \left(1-\frac{2M}{r} +\frac{Q^2}{r^2}\right)} -r^2(d \theta^2  \\
+ \sin^2 \theta d\phi^2). 
\end{multline}
We also note that when $Q=0$, it boils down to ordinary Schwarzschild solutions.

\subsection{Ayon-Beato-Garcia (ABG)  metric}

We note that for non-linear Maxwell's equation, the energy-momentum tensor is given by, 
\begin{equation}\label{eq:43}
    T_{\mu \nu}= \mathcal{L} (F)g_{\mu \nu}-\frac{d \mathcal{L}(F)}{dF} F_{\mu \sigma} F^{\sigma}_{\nu}.
\end{equation}
Here, we have taken a standard ansatz based on \cite{Bronnikov/2001, Matyjasek/2013}, which would satisfy Einstein's equation.

\begin{equation}\label{eq:44}
    \mathcal{H}(P)=P\left[   1-\tanh^2 \left(\frac{|Q|}{2M}(-2Q^2P)^{\frac{1}{4}}\right)\right],
\end{equation}

where $P$ is defined as $P=F\left(\frac{\partial \mathcal{L}}{\partial F}\right)^2$ and $\mathcal{H}(P)$ is obtained from the Legendre transformation as $\mathcal{H}= 2F \frac{\partial \mathcal{L}}{\partial F} - \mathcal{L}$ with $Q$ and $M$ are the charge and mass of the
BH respectively.

\begin{equation}
    ds^2=f(r) dt^2 - \frac{1}{f(r)} dr^2 - r^2(d\theta^2 + \sin^{2} \theta \, d\phi^2),
\end{equation}

where

\begin{equation}
    f(r)=1-\frac{2M}{r} + \frac{2M}{r} \tanh\left(\frac{Q^2}{2Mr}\right).
\end{equation}
We would like to note that as $\lim\limits_{x \to 0} \frac{tanh x}{x} = 1$ so at low $Q$ or high $r$ $f(r)$ coincides with R-N solution. \\
We note that such a metric has already been used in the context of thin-shell wormhole \cite{Rahman/2021, Eloy/1999}.

\subsection{Charged Bardeen black hole}
We note that the Bardeen metric is given by:
\begin{multline*}
    ds^2=\left(1-\frac{2Mr^2}{(r^2+Q^2)^\frac{3}{2}}\right)dt^2- \frac{dr^2}{\left(1-\frac{2Mr^2}{(r^2+Q^2)^\frac{3}{2}}\right)}\\
    -r^2(d\theta^2+sin^2\theta d \phi^2).
\end{multline*}

This was motivated by the fact that it is the regular solution of a collapsing scenario in case some residue charge is there \cite{Bardeen/1968}. \\
As it is clear from the Kretschmann scalar:
$\lim _{r\rightarrow 0}R^{\alpha \beta \gamma \delta }R_{\alpha \beta \gamma \delta }=\frac{96M^2}{Q^{8/3}}$, 
 that it is regular at the center.\\
Now, as discussed in \cite{Ramadhan/2023}, one can take the extreme of two horizons for the metric and can get the Bardeen metric in the presence of non-linear Maxwell's equations as follows:
\begin{equation} 
 f(r)=1 - \frac{2Mr^2}{(r^2 + Q^2)^\frac{3}{2}}+\frac{Q^2 r^2}{\left( r+2+Q^2\right) ^2}. 
\end{equation}
It is also worth noting that this Bardeen metric can also be derived from the mono pole solution of non-linear Maxwell's equation as shown by E. Ayon-Beato, A. Garcia \cite{Ayon-Beato/2000}. So, we can say that this modified Bardeen metric is indeed compatible with the non-linear Maxwell's equations as well.
\section{JUNCTION CONDITION} \label{sec:VII}
As we mentioned before, in order for gravastar to mimic the black hole entropy, one has to implement the fact that it has a shell made out of stiff matter ($\omega=1$). We also note that in order for the solution to be compatible with the external region, we have to invoke the Israel junction condition.\\
 The boundary value problems in GR are sufficiently old and were first studied by Sen \cite{Sen/1924}  in 1924. It was further extended by Lanczos \cite{Lanczos/1924}  by noting some of the subtle difficulties it poses due to choosing a preferred coordinate system. The first complete and general covariance preservation formulation was given by Darmois \cite{Darmois/1927}. However, in this paper, we will follow the approach developed by Israel \cite{Israel/1967, Israel/1966}, which incorporates the general boundary condition in terms of distribution as well as gives the recipe to calculate the physical quantities such as surface pressure ($p$) and surface energy density ($\zeta$) in terms of the metrics.\\
 Here, we offer a very brief account of how the surface pressure ($p$) and energy density ($\zeta$) are calculated. \\
 We first note that for all practical purposes, we can take the thin shell as a zero thickness object so we can split the entire gravastar manifolds into three distinct regions that are $\mathcal{M}_+$ for outside and $\mathcal{M}_-$ for interior and $\Sigma$ for the shell. So the total manifold of gravastar becomes $\mathcal{M}_+ \cup \Sigma \cup \mathcal{M}_-$. We also note that let the $g_{\mu \nu}^{+}(x^{\mu}_{+})$ and $g_{\mu \nu}^{-}(x^{\mu}_{})$ are the metric for the external and interior manifolds respectively. It can be shown (from the killing equation of the manifold) \cite{Israel/1967, Israel/1966} that the shell should satisfy an FRLW metric (2+1 dimensional) like:
 \begin{equation} 
    ds^2=-d\tau^2 + \textbf{a}(t) d\Omega^2.
\end{equation}
 We also like to note that the (2+1) dimensional stress-energy tensor $S_{ij}$ is defined as:
 \begin{equation}
    S_{ij}=-\frac{1}{8 \pi}(k_{ij}-\delta_{ij} k_{\gamma \gamma}) .
\end{equation}
Where $k_{ij}$ (extrinsic curvature or second fundamental form) is defined with respect to the embedding of the manifold by the following formula: 
\begin{equation}
     K^{\pm}_{ij}= -n^{\pm}_{\nu}\left(\frac{\partial^2 x^{\nu}}{\partial \phi^{i} \partial \phi^{j}}+ \Gamma^l_{km} \frac{\partial x^l}{\partial \phi^i} \frac{\partial x^m}{\partial \phi^j} \right).
\end{equation}
Where  $n^{\pm}$ is the normal perpendicular to the surface defined as:
\begin{equation}
    n^{\pm}=\pm \left|g^{lm} \frac{\partial f}{\partial x^{l}} \frac{\partial f}{\partial x^{m}} \right|^{-1/2} \frac{\partial f}{\partial x^{\nu}}, 
\end{equation}
and $\phi$ stands for the intrinsic coordinate, which comes from the embedding of the shell. \\
We would also like to note that the total energy-momentum tensor for the full spacetime can be written as $T_{\mu\nu}$ as $T_{\mu\nu}=\Theta(l)T_{\mu \nu}^++\Theta(-l)T_{\mu \nu}^- +\delta(l)S_{\mu\nu}$, where $\Theta$ is the Heaviside step function and $\delta$ is the Dirac-delta function. \\
As in $S_{ij}=diag(-\varsigma, P)$, we can equate the thin-shell energy density and pressure by the following formula:

\begin{equation} 
     \varsigma=-\frac{1}{4 \pi \textbf{a} }\left[\sqrt {f}\right]^{+}_{-}, 
\end{equation}
and
\begin{equation}
         p=-\frac{\varsigma}{2}+\frac{1}{16 \pi}\left[\frac{f^{\prime}}{\sqrt {f}}\right]^{+}_{-} .
\end{equation}
The aforementioned equations can be used to derive the expressions for the previously mentioned quantities.
By following the recommendations made in  \cite{Forghani/2020,MSharif}, we can observe that. Additionally, by noting the conservation of momentum and energy, we can compute the potential $V(\textbf{a})$ as follows:
\begin{equation} 
    \frac{d}{d\tau}(\varsigma \phi)=p\frac{d\phi}{d\tau}=0\,,
\end{equation}
therefore, $\phi = 4\pi \textbf{a}^2$. Using the conservation equation above, we can ascertain

\begin{equation}
    \varsigma '=-\frac{2}{\textbf{a}}(\varsigma+p)\,.
\end{equation}

According to the prescription in \cite{eric}, the final equation takes the form $\dot{\textbf{a}}^2+V(\textbf{a})$. Consequently, we can obtain, using the equation above:

\begin{equation}\label{eq:56.1}
    V(\textbf{a})=\frac{f(\textbf{a})}{2}+\frac{F(\textbf{a})}{2}-\frac{(f(\textbf{a})-F(\textbf{a}))^2}{64\textbf{a}^2\pi^2\varsigma^2}-4\textbf{a}^2\pi^2\varsigma^2. 
\end{equation}
We would like to note that in Fig-\ref{fig-5} we have plotted all the potentials for various values of $Q$. Also, it is worth noting that there is a unique minima in all of the above cases so keeps on changing as a function of $Q$, so in principle one can use this fact to find the ISCO (Inner Most Stable Circular Orbit) for the gravastar. Which later can be used for phenomenological observations of the accretion disks around gravastars.
\subsection{R-N metric}
We note that for the noncommutative geometry under the charge $Q$, the metric is already well known and given by \cite{Ovgun/2017}, 
\begin{multline}
f(r)_=1-\frac{4M}{r \sqrt{\pi}} \gamma\left(\frac{3}{2},\frac{r^2}{4 \alpha}\right)+\frac{Q^2}{r^2 \pi}\left[\gamma^2\left(\frac{1}{2},\frac{r^2}{4 \alpha}\right) \right.\\ \left.-
\frac{r}{\sqrt{2 \alpha}} \gamma\left(\frac{1}{2},\frac{r^2}{2 \alpha}\right)\right],
\end{multline}

\begin{equation}
    f(r)_+=1 -\frac{2M}{r}+\frac{Q^2}{r^2}.
\end{equation}


\begin{equation}
     \varsigma=-\frac{\iota_1 - \iota_2}{4\pi\textbf{a}}.
\end{equation}

\begin{multline}
    p=\frac{\iota_1 -\iota_2}{8 \pi \textbf{a}}+\frac{1}{16 \pi} \left( \frac{\frac{2M}{\textbf{a}^2}-\frac{2Q^2}{\textbf{a}^3}}{\iota_1}  -\frac{1}{2\pi\textbf{a}^3\alpha \iota_2} \left(4 \sqrt{\pi } \sqrt{\alpha } \right.\right.\\ \left.\left.\text{erf}\left(\frac{\textbf{a}}{2 \sqrt{\alpha }}\right) \left(\sqrt{\pi } \alpha  M \sqrt{\frac{\textbf{a}^2}{\alpha }}+\textbf{a} Q^2 e^{-\frac{\textbf{a}^2}{4 \alpha }}\right) +\sqrt{2 \pi } \textbf{a} \sqrt{\alpha } Q^2 \right.\right. \\\left.\left.\text{erf}\left(\frac{\sqrt{\frac{\textbf{a}^2}{\alpha }}}{\sqrt{2}}\right) - 2 \textbf{a} e^{-\frac{\textbf{a}^2}{2 \alpha }} \sqrt{\frac{\textbf{a}^2}{\alpha }} \left(\sqrt{\pi } M e^{\frac{\textbf{a}^2}{4 \alpha }} \left(\textbf{a}^2+2 \alpha \right)+\sqrt{\alpha } Q^2\right)\right.\right. \\\left.\left. - 4 \pi  \alpha  Q^2 \text{erf}\left(\frac{\textbf{a}}{2 \sqrt{\alpha }}\right)^2\right)       \right).
\end{multline}

\subsection{ABG metric}

\begin{multline}
    f(r)_{-}=1-\frac{4M}{r \sqrt{\pi}} \gamma\left(\frac{3}{2},\frac{r^2}{4 \alpha}\right)+\frac{Q^2}{r^2 \pi}\left[\gamma^2\left(\frac{1}{2},\frac{r^2}{4 \alpha}\right) \right.\\ \left.
    -\frac{r}{\sqrt{2 \alpha}} \gamma\left(\frac{1}{2},\frac{r^2}{2 \alpha}\right)\right],
\end{multline}

\begin{equation}
    f(r)_+=1-\frac{2M}{r}+ \frac{2M}{r}\tanh \left(\frac{Q^2}{2Mr}\right).
\end{equation}


\begin{equation}
    \varsigma=-\frac{\iota_3-\iota_2}{4\pi\textbf{a}}.
\end{equation}

\begin{multline}
    p=\frac{\frac{4 \textbf{a} M}{e^{\frac{Q^2}{\textbf{a} M}}+1}-Q^2 \text{sech}^2\left(\frac{Q^2}{2 \textbf{a} M}\right)}{16 \pi  \textbf{a}^3 \iota_3 }+\frac{\iota_3-\iota_2}{8\pi\textbf{a}}-\frac{1}{32\pi ^2 \textbf{a}^3 \alpha \iota_2}\\
    \left(4 \sqrt{\pi } \sqrt{\alpha } \text{erf}\left(\frac{\textbf{a}}{2 \sqrt{\alpha }}\right) \left(\sqrt{\pi } \alpha  M \sqrt{\frac{\textbf{a}^2}{\alpha }}+\textbf{a} Q^2 e^{-\frac{\textbf{a}^2}{4 \alpha }}\right) \right. \\ \left.
        +\sqrt{2 \pi } \textbf{a} \sqrt{\alpha } Q^2 \text{erf}\left(\frac{\sqrt{\frac{\textbf{a}^2}{\alpha }}}{\sqrt{2}}\right)-2 \textbf{a} e^{-\frac{\textbf{a}^2}{2 \alpha }} \sqrt{\frac{\textbf{a}^2}{\alpha }} \left(\sqrt{\pi } M e^{\frac{\textbf{a}^2}{4 \alpha }} \right.\right.\\\left.\left.\left(\textbf{a}^2+2 \alpha \right)+\sqrt{\alpha } Q^2\right)-4 \pi  \alpha  Q^2 \text{erf}\left(\frac{\textbf{a}}{2 \sqrt{\alpha }}\right)^2\right).
\end{multline}


\subsection{Charged Bardeen black hole}

\begin{multline}
    f(r)_{-}= 1-\frac{4M}{r \sqrt{\pi}} \gamma\left(\frac{3}{2},\frac{r^2}{4 \alpha}\right)+\frac{Q^2}{r^2 \pi}\left[\gamma^2\left(\frac{1}{2},\frac{r^2}{4 \alpha}\right) \right.\\ \left.
    -\frac{r}{\sqrt{2 \alpha}} \gamma\left(\frac{1}{2},\frac{r^2}{2 \alpha}\right)\right],
\end{multline}

\begin{equation}
    f(r)_+=1-\frac{2Mr^2}{(r^2+Q^2)^{\frac{3}{2}}}+\frac{Q^2 r^2}{(r+2+Q^2)^2} .
\end{equation}


\begin{equation}
    \varsigma=-\frac{\iota_4-\iota_2}{4\pi\textbf{a}}.
\end{equation}

\begin{multline}
    p=\frac{\iota_4-\iota_2}{8\pi\textbf{a}}+\frac{1}{16 \pi} \left( \frac{1}{\iota_4} \left( -\frac{4 \textbf{a} M \left(\sqrt{\pi }-2 \Gamma \left(\frac{3}{2},\frac{\textbf{a}^2}{4 \alpha }\right)\right)}{\sqrt{\pi } \left(\textbf{a}^2+Q^2\right)^{3/2}} \right.\right.\\ \left.\left.-\frac{\textbf{a}^2 Q^2}{\left(\textbf{a}+Q^2+2\right)^2}-\frac{\textbf{a}^3 M e^{-\frac{\textbf{a}^2}{4 \alpha }} \sqrt{\frac{\textbf{a}^2}{\alpha }}}{\sqrt{\pi } \alpha  \left(\textbf{a}^2+Q^2\right)^{3/2}}\right.\right.\\\left.\left.+ \frac{6 \textbf{a}^3 M \left(\sqrt{\pi }-2 \Gamma \left(\frac{3}{2},\frac{\textbf{a}^2}{4 \alpha }\right)\right)}{\sqrt{\pi } \left(\textbf{a}^2+Q^2\right)^{5/2}}+\frac{2 \textbf{a} Q^2}{\textbf{a}+Q^2+2} \right)\right.\\\left.
    -\frac{1}{2 \pi  \textbf{a}^3 \alpha \iota_2}
      \left(4 \sqrt{\pi } \sqrt{\alpha } \text{erf}\left(\frac{\textbf{a}}{2 \sqrt{\alpha }}\right) \left(\sqrt{\pi } \alpha  M \sqrt{\frac{\textbf{a}^2}{\alpha }}+\textbf{a} Q^2 e^{-\frac{\textbf{a}^2}{4 \alpha }}\right) \right.\right. \\ \left. \left. 
        +\sqrt{2 \pi } \textbf{a} \sqrt{\alpha } Q^2 \text{erf}\left(\frac{\sqrt{\frac{\textbf{a}^2}{\alpha }}}{\sqrt{2}}\right)-2 \textbf{a} e^{-\frac{\textbf{a}^2}{2 \alpha }} \sqrt{\frac{\textbf{a}^2}{\alpha }} \left(\sqrt{\pi } M e^{\frac{\textbf{a}^2}{4 \alpha }} \left(\textbf{a}^2+2 \alpha \right) \right.\right.\right.\\ \left.\left.\left.
        +\sqrt{\alpha } Q^2\right)-4 \pi  \alpha  Q^2 \text{erf}\left(\frac{\textbf{a}}{2 \sqrt{\alpha }}\right)^2\right)        \right).
\end{multline}

Where,
\begin{multline}\label{eq:66}
    \iota_1=\sqrt{\frac{\textbf{a}^2-2 \textbf{a} M+Q^2}{\textbf{a}^2}},\\
     \iota_2=\left(   \frac{\frac{4 M \Gamma \left(\frac{3}{2},\frac{\textbf{a}^2}{4 \alpha }\right)}{\sqrt{\pi }}+Q^2 \text{erf}\left(\frac{\textbf{a}}{2 \sqrt{\alpha }}\right)^2}{\textbf{a}^2} -\frac{Q^2 \text{erf}\left(\frac{\sqrt{\frac{\textbf{a}^2}{\alpha }}}{\sqrt{2}}\right)}{\sqrt{2 \pi } \textbf{a} \sqrt{\alpha }}  -\frac{2M}{\textbf{a}}+1     \right)^{\frac{1}{2}}  , \\
    \iota_3=\sqrt{\frac{2 M \tanh \left(\frac{Q^2}{2 \textbf{a} M}\right)}{\textbf{a}}-\frac{2 M}{\textbf{a}}+1},\\
    \iota_4=\sqrt{-\frac{2\textbf{a}^2 M \left(\sqrt{\pi }-2 \Gamma \left(\frac{3}{2},\frac{\textbf{a}^2}{4 \alpha }\right)\right)}{\sqrt{\pi } \left(\textbf{a}^2+Q^2\right)^{3/2}}+\frac{\textbf{a}^2 Q^2}{\textbf{a}+Q^2+2}+1}.
\end{multline}


\begin{widetext}

    \begin{figure}[]
        \centering
       \subfigure[]{\includegraphics[width=7cm,height=5cm]{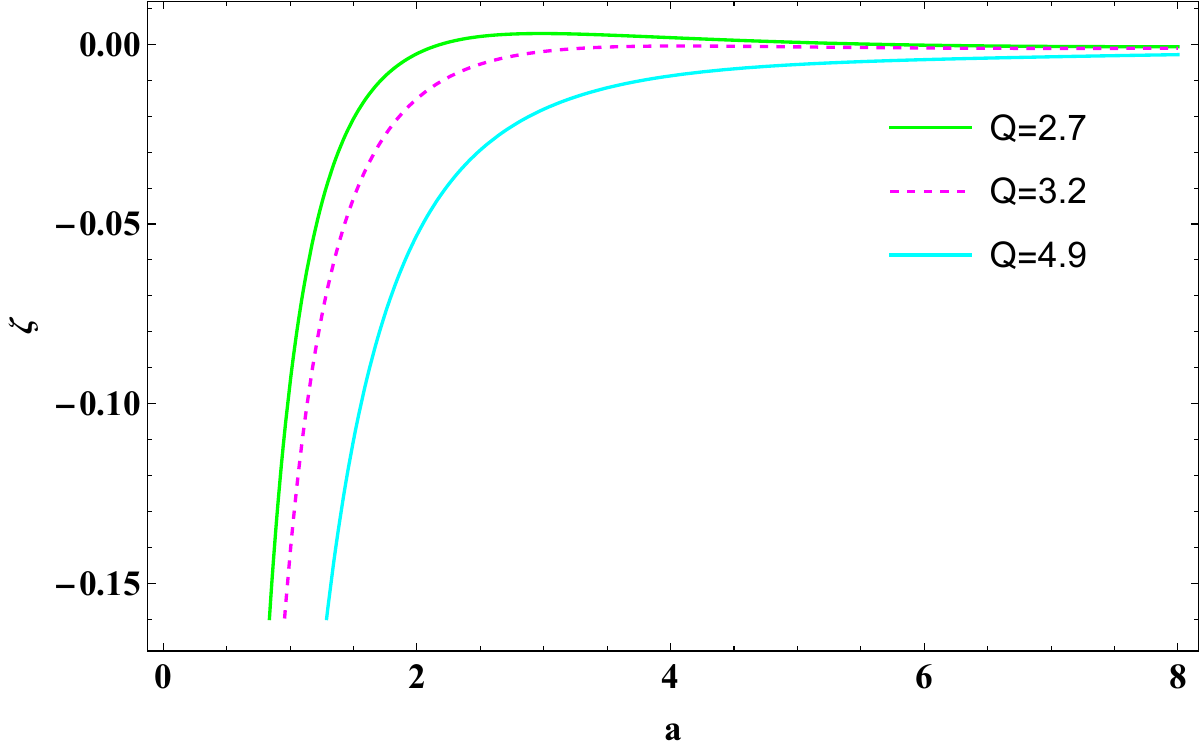}}\,\,\,\,\,\,\,\,\,\,\,\,\,\,\,\,\,\,\,\,\,\,\,\,\,\,
       \subfigure[]{\includegraphics[width=7cm,height=5cm]{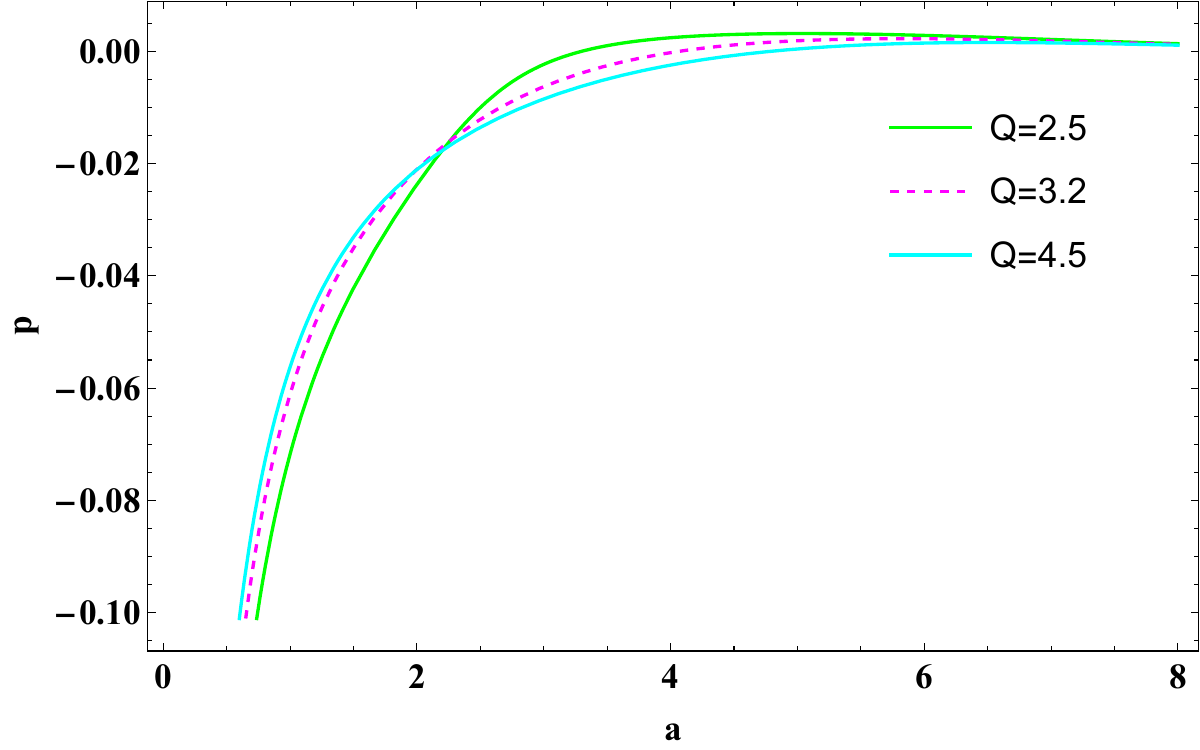}}\,\,\,\,\,\,\,\,\,\,\,\,\,\,\,\,\,\,\,\,\,
       \subfigure[]{\includegraphics[width=7cm,height=5cm]{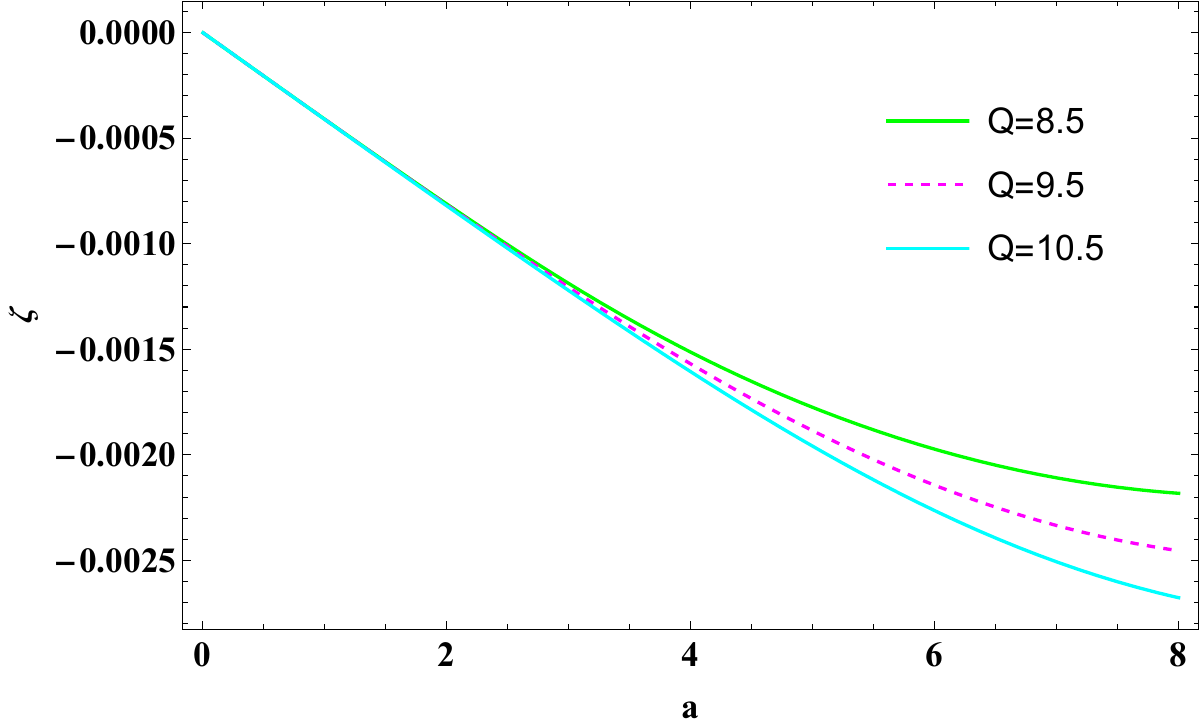}}\,\,\,\,\,\,\,\,\,\,\,\,\,\,\,\,\,\,\,\,\,
       \subfigure[]{\includegraphics[width=7cm,height=5cm]{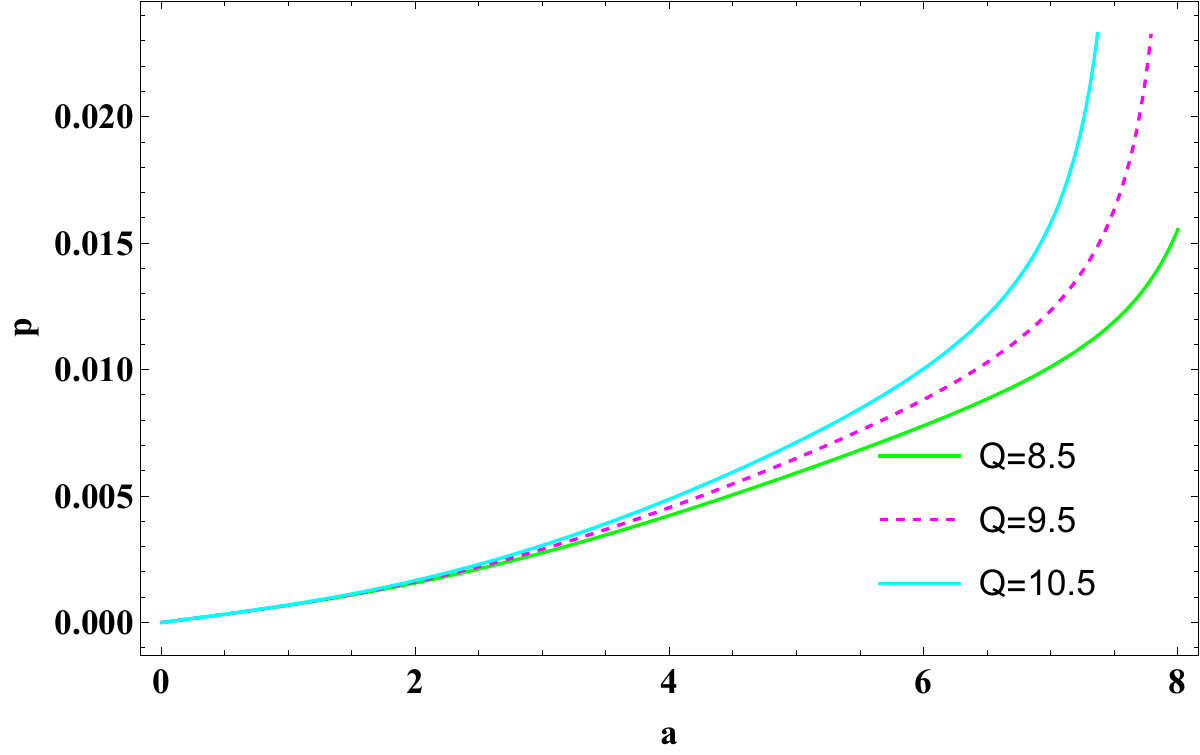}}\,\,\,\,\,\,\,\,\,\,\,\,\,\,\,\,\,\,\,\,\,
        \subfigure[]{\includegraphics[width=7cm,height=5cm]{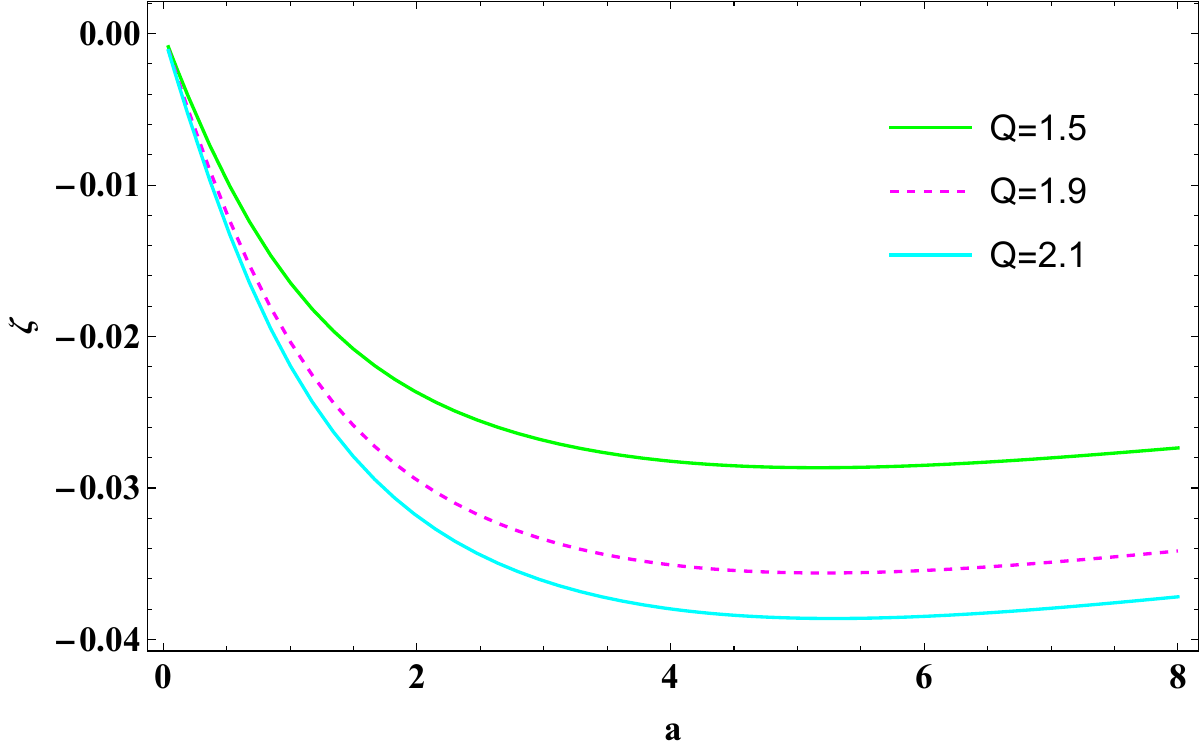}}\,\,\,\,\,\,\,\,\,\,\,\,\,\,\,\,\,\,\,\,\,
        \subfigure[]{\includegraphics[width=7cm,height=5cm]{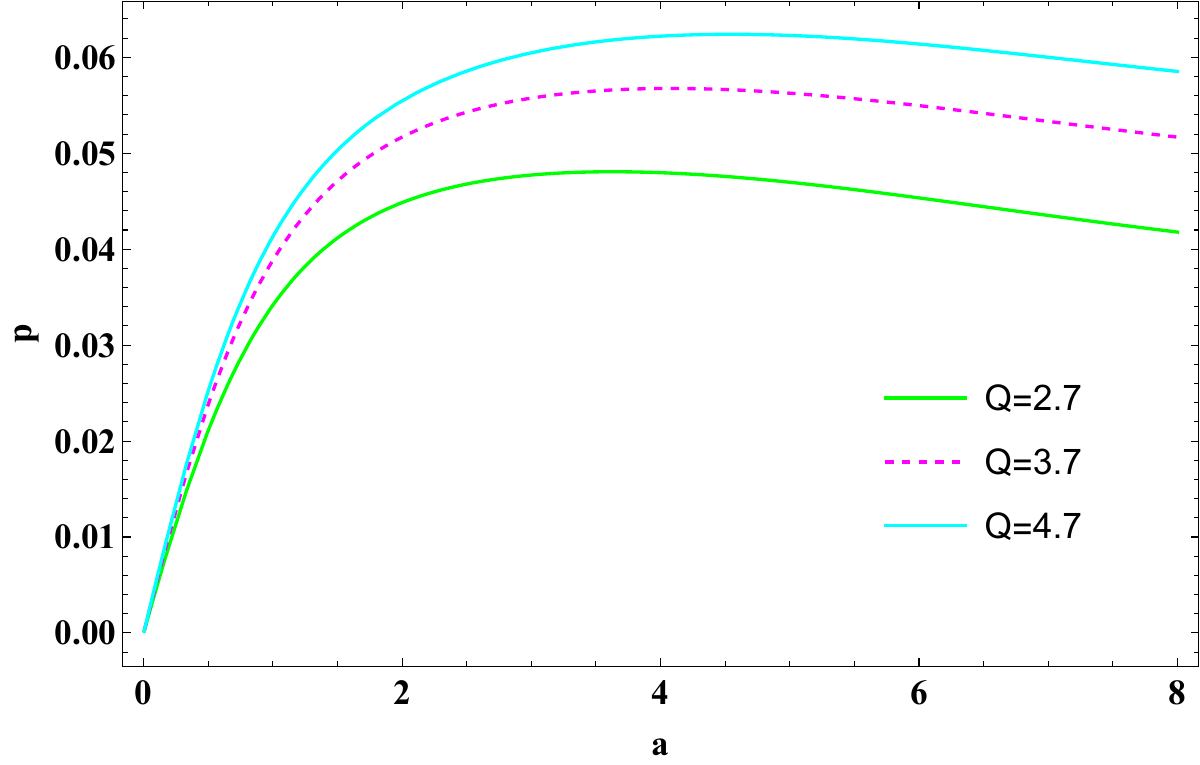}}\,\,\,\,\,\,\,\,\,\,\,\,\,\,\,\,\,\,\,\,\,
        \caption{ The figure illustrates the profile of ,(a) Variation of surface energy density inside the thin shell in R-N metric, (b) variation of surface pressure inside the thin shell in R-N metric, (c) Variation of surface energy density inside the thin shell in ABG metric, (d) Variation of surface pressure inside the thin shell in ABG metric, (e) Variation of surface energy density inside the thin shell in charged Bardeen metric, and (f)  variation of surface pressure inside the thin shell in charged Bardeen metric}
        \label{fig-4}
    \end{figure}
    
\end{widetext}






\begin{widetext}

       \begin{figure}[h]
        \centering
       \subfigure[]{\includegraphics[width=7cm,height=5cm]{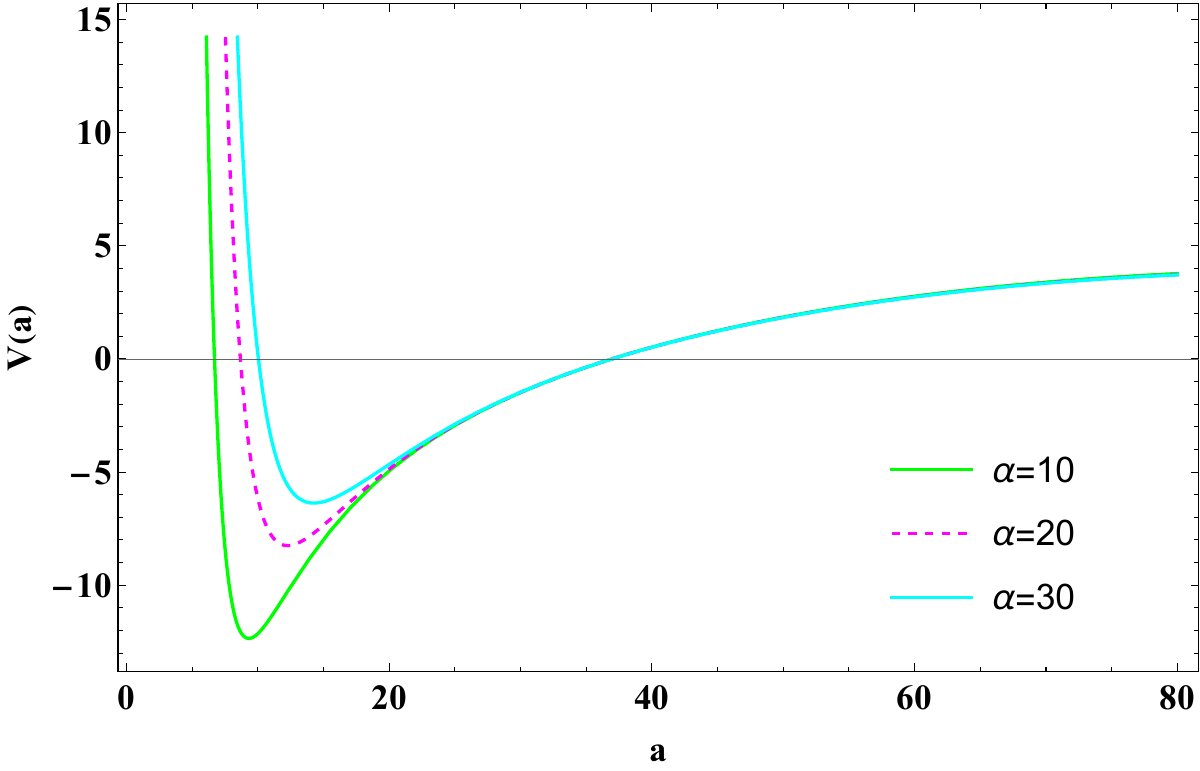}}\,\,\,\,\,\,\,\,\,\,\,\,\,\,\,\,\,\,\,\,\,\,\,\,\,\,
       \subfigure[]{\includegraphics[width=7cm,height=5cm]{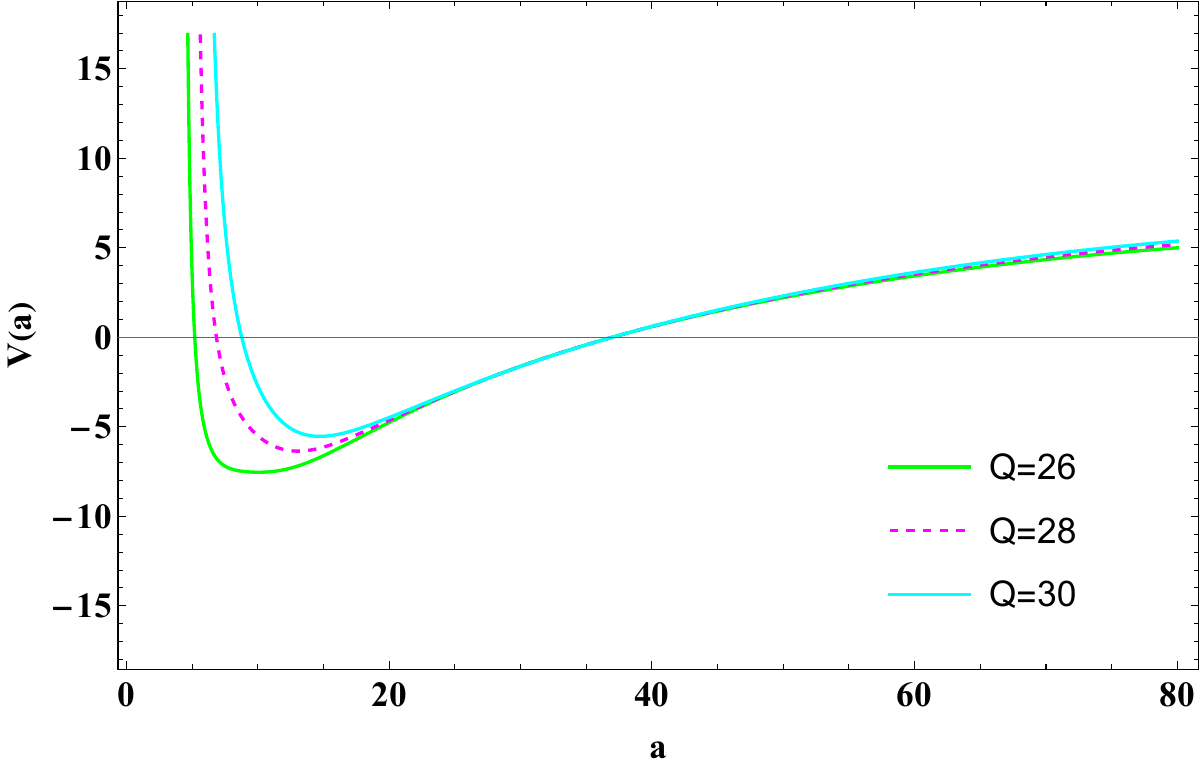}}\,\,\,\,\,\,\,\,\,\,\,\,\,\,\,\,\,\,\,\,\,
       \subfigure[]{\includegraphics[width=7cm,height=5cm]{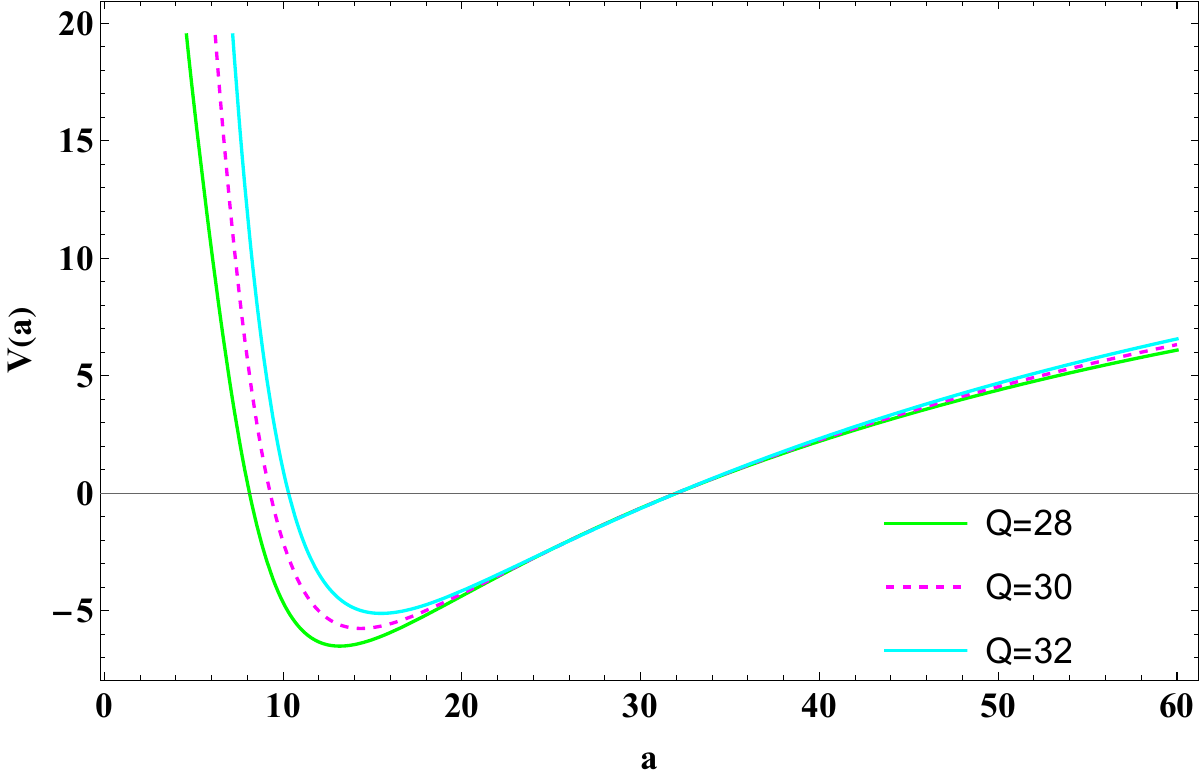}}\,\,\,\,\,\,\,\,\,\,\,\,\,\,\,\,\,\,\,\,\,
       \subfigure[]{\includegraphics[width=7cm,height=5cm]{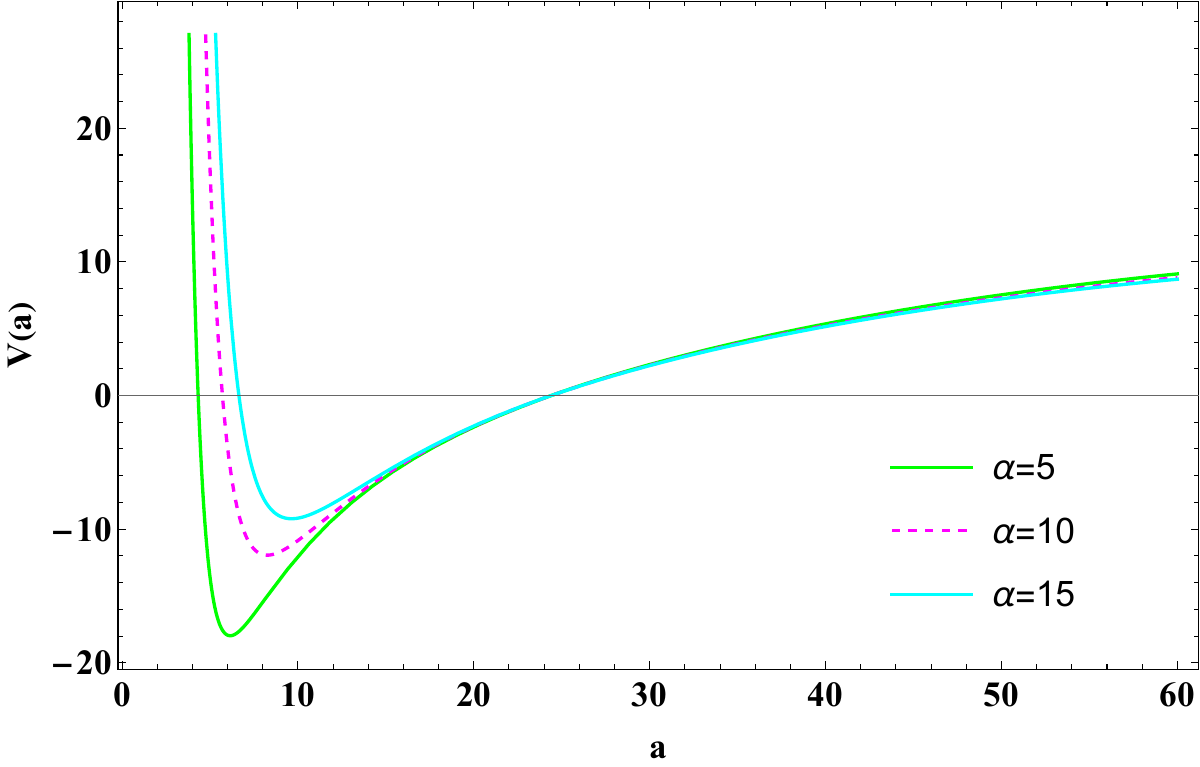}}\,\,\,\,\,\,\,\,\,\,\,\,\,\,\,\,\,\,\,\,\,
        \subfigure[]{\includegraphics[width=7cm,height=5cm]{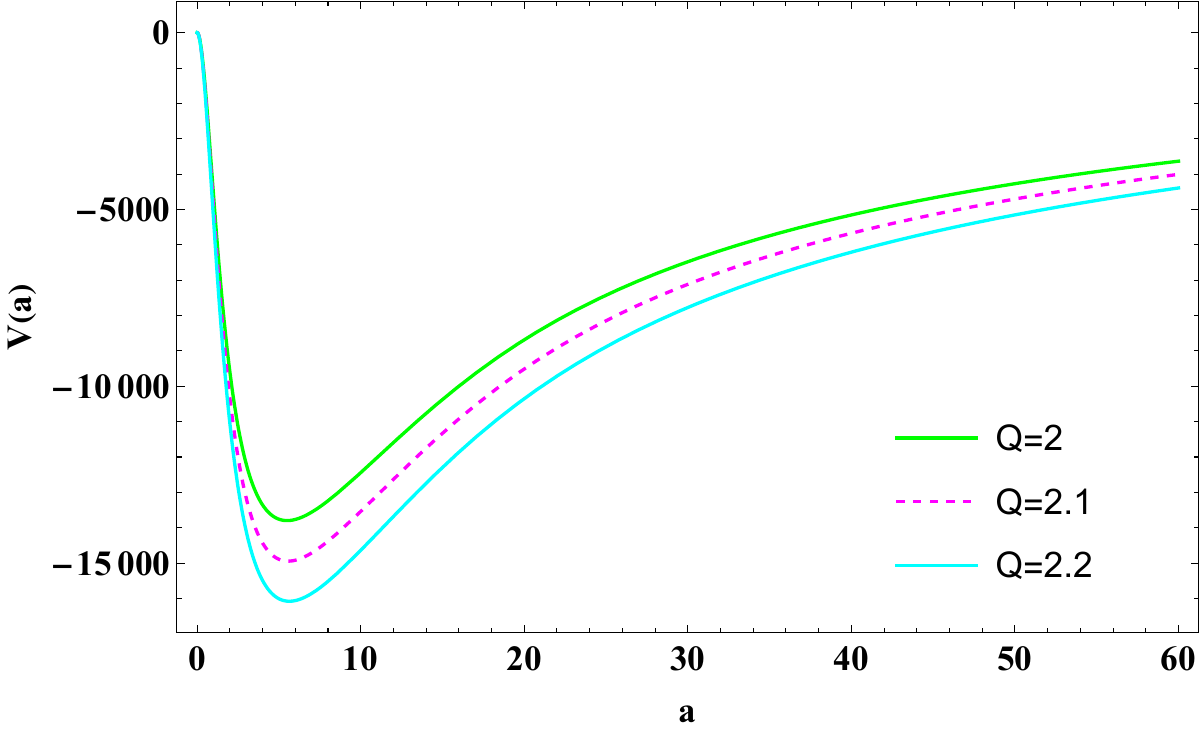}}\,\,\,\,\,\,\,\,\,\,\,\,\,\,\,\,\,\,\,\,\,
        \subfigure[]{\includegraphics[width=7cm,height=5cm]{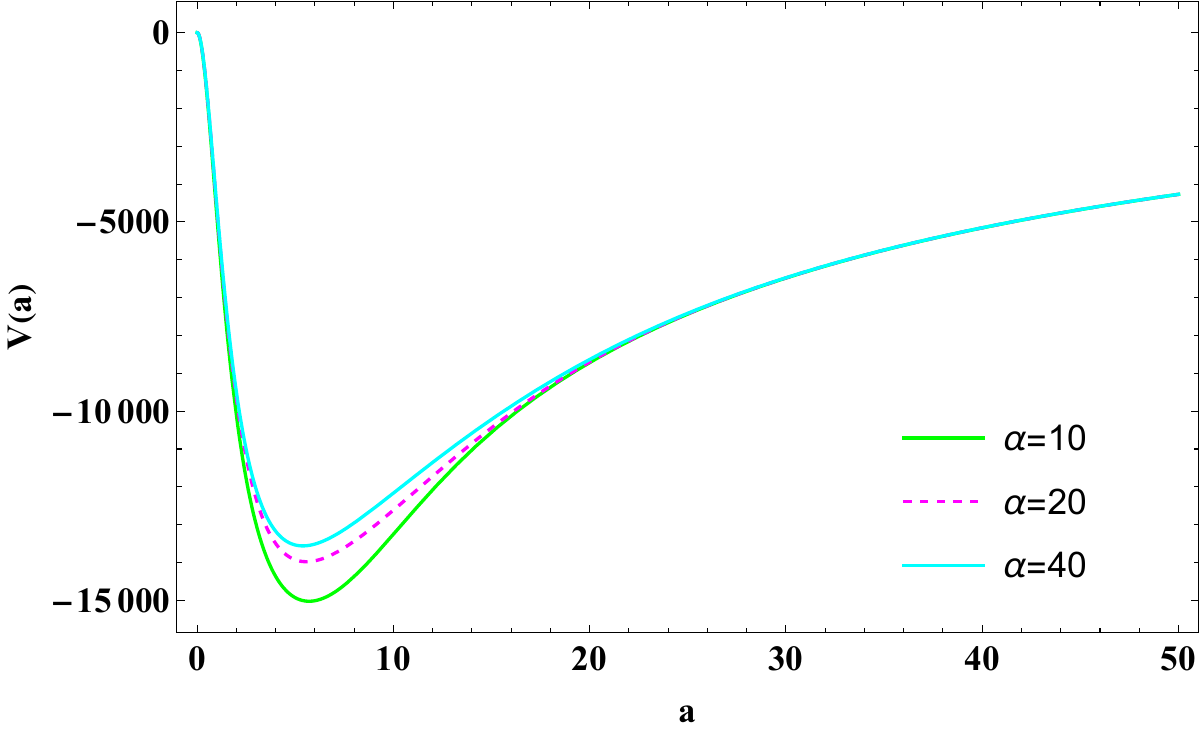}}\,\,\,\,\,\,\,\,\,\,\,\,\,\,\,\,\,\,\,\,\,
        \caption{The figure illustrates the profile of Potential for the shell: (a) outside the Reissner-Nordstrom (R-N) metric for different values of $\alpha$, (b) outside the R-N metric for different values of $Q$, (c) outside the Ayon-Beato-Garcia (ABG) metric for different values of $Q$, (d) outside the ABG metric for different values of $\alpha$, (e) outside the charged Bardeen metric for different values of $Q$, and (f) outside the charged Bardeen metric for different values of $\alpha$.}
        \label{fig-5}
    \end{figure}

\end{widetext}

\section{Physical features of the model} \label{sec:VIII}

\subsection{Proper length} \label{subsec:A}

According to the theories of Mazur and Mottola \cite{P/2023, Mazur/2004}, the shell is located between the intersection of two space-times. Between the inner region and the intermediate thin shell, the phase barrier $r_1 = d $ and the phase border $r_2=d+\epsilon$, which separates the outside space-time from the intermediate thin shell, are where the shell's length changes. The following formula can be used to determine the necessary length or appropriate thickness of the shell as well as the proper thickness between these two interfaces:

\begin{equation} \label{eq:67}  
 \begin{gathered} 
     \textit{l}=\int^{d+\epsilon}_{d} \sqrt{e^{\lambda(r)}} dr =\int^{d+\epsilon}_{d} \frac{1}{\sqrt{1+\frac{b \left(r^4-16 c_1\right)}{4 a r^2}}} dr \\ = \int^{d+\epsilon}_{d} \frac{1}{f(r)},
\end{gathered}
\end{equation}

where $f(r)=\sqrt{1+\frac{b \left(r^4-16 c_1\right)}{4 a r^2}}$.
Currently, it is not easy to evaluate the integral given in equation \eqref{eq:67}. To solve the previous integral, let us select $\frac{d f(r)}{dr}=\frac{1}{f(r)}$. As a result, we get,

\begin{equation} \label{eq:68}
    \textit{l}= f(d+\epsilon)-f(d). 
\end{equation}

Keeping $\epsilon$'s linear order, we derive from equation \eqref{eq:68} after expanding $f(d+\epsilon)$ in the Taylor series around `$d$'.

\begin{equation} \label{eq:69}
      \textit{l}= \epsilon \frac{d f(r)}{dr} \approx \frac{\epsilon}{\sqrt{1+\frac{b \left(r^4-16 c_1\right)}{4 a r^2}}}.
\end{equation}

Due to the exceptionally tiny value of $\epsilon$, the exponential higher-order terms can be disregarded. The appropriate variation in length for the radius of the thin shell is illustrated in Fig. \eqref{fig-6}(a), which shows how it is monotonically increasing.


\subsection{Entropy}\label{subsec:B}

The core region of a gravastar contains no entropy density, suggesting a stable state for a singular condensate region. As per the research done by Mazur and Mottola, the entropy present in the gravastar's thin middle layer can be determined through a specific formula \cite{P/2023, Mazur/2004}.

\begin{equation} \label{eq:70}
   \mathcal{S}= \int^{d+\epsilon}_{d} 4 \pi r^2 s(r) \sqrt{e^{\lambda}} dr.
\end{equation}

The entropy of the thin shell can be determined in the state equation, $p = \rho = \frac{k^2}{8 \pi G}$, where $G$ is included for dimensional reasons, therefore ${k}^2$ is a dimensionless constant. A relativistic fluid with zero chemical potential has a local specific entropy density of $sT = p + \rho$, according to the standard thermodynamic Gibbs relation \cite{P/2023}.
\begin{equation}\label{eq:71}
    s(r)= \frac{k^2 {K_B}^2 T(r)}{4 \pi \hbar^2 G}=\frac{k K_B}{\hbar}\left(\frac{p}{2 \pi G}\right)^{\frac{1}{2}}.
\end{equation}

Here, we have taken $G=1$. So Eq.\eqref{eq:71} becomes 
\begin{multline} \label{eq:72}
\begin{gathered}
    S= \int^{d+\epsilon}_{d} 4 \pi r^2 k \left( \frac{K_B}{\hbar}\right) \left(\frac{p}{2 \pi}\right)^{\frac{1}{2}} \sqrt{e^{\lambda}} dr \\ =\int^{d+\epsilon}_{d} 2 \sqrt{2 \pi} r^2 k \left( \frac{K_B} {\hbar}\right) \sqrt{p e^{\lambda}} dr .
    \end{gathered}
\end{multline}

Now Eq.\eqref{eq:72} can also be written as

\begin{equation}\label{eq:73}
    S= 2 \sqrt{2 \pi}r^2 k \left(  \frac{K_B}{\hbar}\right) N,
\end{equation}

where
\begin{equation}\label{eq:74}
    N=\int^{d+\epsilon}_{d} W(r) dr .
\end{equation}

and

\begin{multline}\label{eq:75}
     W(r)=\frac{1}{\sqrt[4]{\frac{b \left(r^4-16 \text{c1}\right)}{4 a r^2}+1}} \left( \frac{\Upsilon_1-\Upsilon_2}{8 \pi r}+\frac{1}{16 \pi}\right.\\\left.\left(\frac{\frac{2 M}{r^2}-\frac{2 Q^2}{r^3}}{\Upsilon_1} -\frac{1}{2\pi \alpha r^3 \Upsilon_2} \left(4 \sqrt{\pi } \sqrt{\alpha } \text{erf}\left(\frac{r}{2 \sqrt{\alpha }}\right) \right.\right.\right.\\ \left.\left.\left.\left(\sqrt{\pi } \alpha  M \sqrt{\frac{r^2}{\alpha }}+Q^2 r e^{-\frac{r^2}{4 \alpha }}\right)+ \sqrt{2 \pi } \sqrt{\alpha } Q^2 r \text{erf}\left(\frac{\sqrt{\frac{r^2}{\alpha }}}{\sqrt{2}}\right) \right.\right.\right.\\\left.\left.\left.-4 \pi  \alpha  Q^2 \text{erf}\left(\frac{r}{2 \sqrt{\alpha }}\right)^2 -2 r e^{-\frac{r^2}{2 \alpha }} \sqrt{\frac{r^2}{\alpha }} \left(\sqrt{\pi } M e^{\frac{r^2}{4 \alpha }} \left(2 \alpha +r^2\right) \right.\right.\right.\right.\\\left.\left.\left.\left.+\sqrt{\alpha } Q^2\right)\right)   \right)\right)^{\frac{1}{2}}.
\end{multline}

At present, solving the integral outlined in equation \eqref{eq:74} presents a significant challenge. Thus, to tackle the calculation of this integral, let's propose that $F(r)$ serves as the anti-derivative of $W(r)$. By utilizing the basic theorem of integral calculus, this transforms equation \eqref{eq:74} accordingly:

\begin{equation}\label{eq:76}
    N=[F(r)]^{d+\epsilon}_{d}=F(d+\epsilon)-F(d).
\end{equation}

In the Taylor series, we expand $F(d+\epsilon)$ around $d$', maintaining the linear order of $\epsilon$ from equation \eqref{eq:76}. We determine from \eqref{eq:72} that

\begin{multline}\label{eq:77}
      S=2 \sqrt{2 \pi} r^2 k \left(\frac{K_B}{\hbar}\right)  \epsilon \frac{1}{\sqrt[4]{\frac{b \left(r^4-16 \text{c1}\right)}{4 a r^2}+1}}  \left( \frac{\Upsilon_1-\Upsilon_2}{8 \pi r} \right.\\\left.+\frac{1}{16 \pi}\left(\frac{\frac{2 M}{r^2}-\frac{2 Q^2}{r^3}}{\Upsilon_1} -\frac{1}{2\pi \alpha r^3 \Upsilon_2} \left(4 \sqrt{\pi } \sqrt{\alpha } \text{erf}\left(\frac{r}{2 \sqrt{\alpha }}\right) \right.\right.\right.\\ \left.\left.\left.\left(\sqrt{\pi } \alpha  M \sqrt{\frac{r^2}{\alpha }}+Q^2 r e^{-\frac{r^2}{4 \alpha }}\right)+ \sqrt{2 \pi } \sqrt{\alpha } Q^2 r \text{erf}\left(\frac{\sqrt{\frac{r^2}{\alpha }}}{\sqrt{2}}\right) \right.\right.\right.\\\left.\left.\left.-4 \pi  \alpha  Q^2 \text{erf}\left(\frac{r}{2 \sqrt{\alpha }}\right)^2 -2 r e^{-\frac{r^2}{2 \alpha }} \sqrt{\frac{r^2}{\alpha }} \left(\sqrt{\pi } M e^{\frac{r^2}{4 \alpha }} \left(2 \alpha +r^2\right) \right.\right.\right.\right.\\\left.\left.\left.\left.+\sqrt{\alpha } Q^2\right)\right)   \right)\right)^{\frac{1}{2}}.
\end{multline}

Where 
\begin{multline}\label{eq:78}
    \Upsilon_1=\sqrt{-\frac{2 M}{r}+\frac{Q^2}{r^2}+1},\\
     \Upsilon_2= \left(\frac{Q^2 \text{erf}\left(\frac{r}{2 \sqrt{\alpha }}\right)^2+\frac{4 M r \Gamma \left(\frac{3}{2},\frac{r^2}{4 \alpha }\right)}{\sqrt{\pi }}}{r^2}-\frac{Q^2 \text{erf}\left(\frac{\sqrt{\frac{r^2}{\alpha }}}{2}\right)}{\sqrt{2 \pi } \sqrt{\alpha } r}\right. \\ \left.
     -\frac{2 M}{r}+1\right)^{\frac{1}{2}}.
\end{multline}

Here, we also noted that the value of $c_1$ is given in equation \eqref{eq:85}.

Thus, the entropy expression for our proposed model was successfully obtained. Eq.\eqref{eq:77} suggests that $S\approx \mathcal{O}(\epsilon)$ if the thin shell's thickness is $\epsilon<<d$. Fig-\eqref{fig-6}(b) shows the variation of entropy with respect to thickness, which shows that monotonically increasing.


\subsection{Energy}\label{subsec:C}
The shell's energy can be computed using the following formula:

\begin{multline}\label{eq:79}
    E=\int^{d+\epsilon}_d 4\pi r^2 \rho \, dr
    \\=\int^{d+\epsilon}_d 4\pi r^2 \left(  \frac{\Upsilon_1-\Upsilon_2}{8 \pi r}+\frac{1}{16\pi}\left( \frac{\frac{2 M}{r^2}-\frac{2 Q^2}{r^3}}{\Upsilon_1}-\frac{1}{2\pi\alpha r^3\Upsilon_2}\right.\right.\\\left.\left.\left( 4 \sqrt{\pi } \sqrt{\alpha } \text{erf}\left(\frac{r}{2 \sqrt{\alpha }}\right) \left(\sqrt{\pi } \alpha  M \sqrt{\frac{r^2}{\alpha }}+Q^2 r e^{-\frac{r^2}{4 \alpha }}\right) \right.\right.\right.\\\left.\left.\left.+ \sqrt{2 \pi } \sqrt{\alpha } Q^2 r \text{erf}\left(\frac{\sqrt{\frac{r^2}{\alpha }}}{\sqrt{2}}\right) - 4 \pi  \alpha  Q^2 \text{erf}\left(\frac{r}{2 \sqrt{\alpha }}\right)^2 - \right.\right.\right.\\  \left.\left.\left.2 r e^{-\frac{r^2}{2 \alpha }} \sqrt{\frac{r^2}{\alpha }} \left(\sqrt{\pi } M e^{\frac{r^2}{4 \alpha }} \left(2 \alpha +r^2\right)+\sqrt{\alpha } Q^2\right)\right) \right)\right)^{\frac{1}{2}} dr.
\end{multline}

It is not easy to evaluate the above integral, so to solve the above integral, let us choose $h(r)$ as the anti-derivative of $g(r)$. Using the fundamental theorem of integral calculus, the equation \eqref{eq:79} modifies as

\begin{equation} \label{eq:80}
    E=[h(r)]^{d+\epsilon}_d= h(d+\epsilon)-h(d).
\end{equation}
We extend $h(d+\epsilon)$ in the Taylor series around `$d$' and then compute the solution from equation \eqref{eq:80} while maintaining the linear order of $\epsilon$.

\begin{multline}\label{eq:81}
     E=4 \pi r^2 \epsilon \left(  \frac{\Upsilon_1-\Upsilon_2}{8 \pi r}+\frac{1}{16\pi}\left( \frac{\frac{2 M}{r^2}-\frac{2 Q^2}{r^3}}{\Upsilon_1}-\frac{1}{2\pi\alpha r^3\Upsilon_2}\right.\right.\\\left.\left.\left( 4 \sqrt{\pi } \sqrt{\alpha } \text{erf}\left(\frac{r}{2 \sqrt{\alpha }}\right) \left(\sqrt{\pi } \alpha  M \sqrt{\frac{r^2}{\alpha }}+Q^2 r e^{-\frac{r^2}{4 \alpha }}\right) \right.\right.\right.\\\left.\left.\left.+ \sqrt{2 \pi } \sqrt{\alpha } Q^2 r \text{erf}\left(\frac{\sqrt{\frac{r^2}{\alpha }}}{\sqrt{2}}\right) - 4 \pi  \alpha  Q^2 \text{erf}\left(\frac{r}{2 \sqrt{\alpha }}\right)^2 - \right.\right.\right.\\  \left.\left.\left.2 r e^{-\frac{r^2}{2 \alpha }} \sqrt{\frac{r^2}{\alpha }} \left(\sqrt{\pi } M e^{\frac{r^2}{4 \alpha }} \left(2 \alpha +r^2\right)+\sqrt{\alpha } Q^2\right)\right) \right)\right)^{\frac{1}{2}}.
\end{multline}

Where value of $\Upsilon_1,\Upsilon_2$ is given in equation \eqref{eq:78}.

In Fig-\eqref{fig-6}(c), we plot the variation of proper length vs thickness $\epsilon$, which is monotonically increasing.


\subsection{The EoS parameter}\label{subsec:D}
The equation of state's state parameter, $\omega$, can be stated as

\begin{multline}\label{eq:82}
    \frac{p}{\varsigma}=-\frac{1}{2}-\frac{1}{4(\iota_1-\iota_2)}\left(\textbf{a} \left(\frac{2(\textbf{a}M-Q^2)}{\textbf{a}^3\iota_1} -\frac{1}{2\pi\textbf{a}^3 \alpha \iota_2}\right.\right.\\\left.\left.\left( 4 \sqrt{\pi } \sqrt{\alpha } \text{erf}\left(\frac{\textbf{a}}{2 \sqrt{\alpha }}\right) \left(\sqrt{\pi } \alpha  M \sqrt{\frac{\textbf{a}^2}{\alpha }}+\textbf{a} Q^2 e^{-\frac{\textbf{a}^2}{4 \alpha }}\right) \right.\right.\right.\\\left.\left.\left.+\sqrt{2 \pi } \textbf{a} \sqrt{\alpha } Q^2 \text{erf}\left(\frac{\sqrt{\frac{\textbf{a}^2}{\alpha }}}{\sqrt{2}}\right)-2 \textbf{a} e^{-\frac{\textbf{a}^2}{2 \alpha }} \sqrt{\frac{\textbf{a}^2}{\alpha }} \right.\right.\right.\\\left.\left.\left. \left(\sqrt{\pi } M e^{\frac{\textbf{a}^2}{4 \alpha }} \left(\textbf{a}^2+2 \alpha \right)+\sqrt{\alpha } Q^2\right)-4 \pi  \alpha  Q^2 \right.\right.\right.\\\left.\left.\left.\text{erf}\left(\frac{\textbf{a}}{2 \sqrt{\alpha }}\right)^2
 \right) \right)  \right).
\end{multline}

We also noted here that value of $\iota_1,\iota_2$
are given in the equation \eqref{eq:66}.

\begin{widetext}

    \begin{figure}[h]
        \centering
       \subfigure[]{\includegraphics[width=6.5cm,height=5cm]{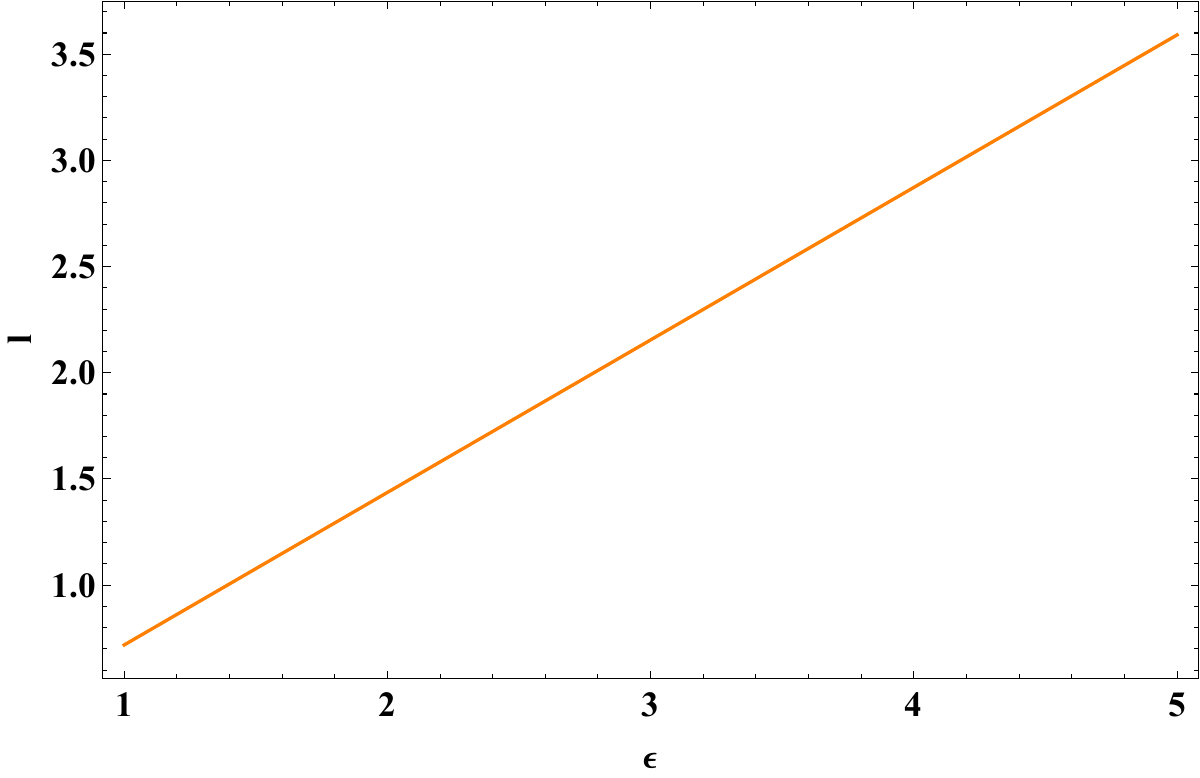}}\,\,\,\,\,\,\,\,\,\,\,\,\,\,\,\,\,\,\,\,\,\,\,\,\,\,
       \subfigure[]{\includegraphics[width=6.5cm,height=5cm]{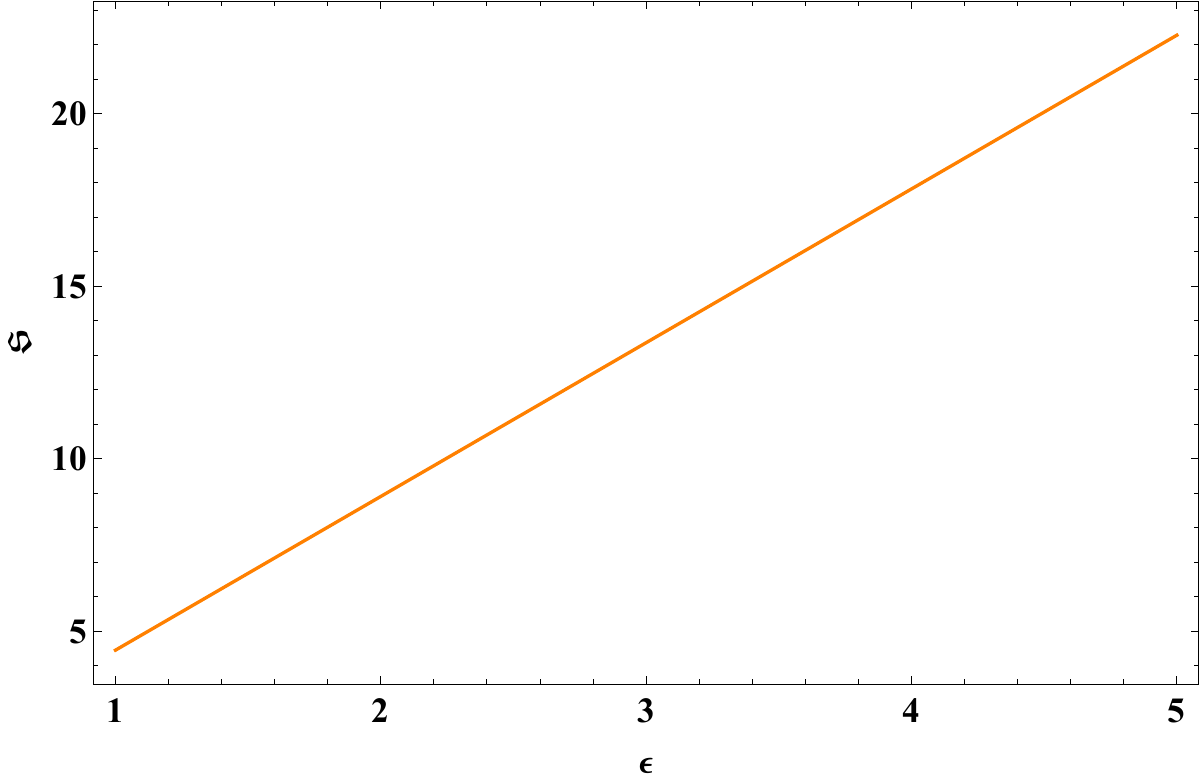}}\,\,\,\,\,\,\,\,\,\,\,\,\,\,\,\,\,\,\,\,\,
       \subfigure[]{\includegraphics[width=6.5cm,height=5cm]{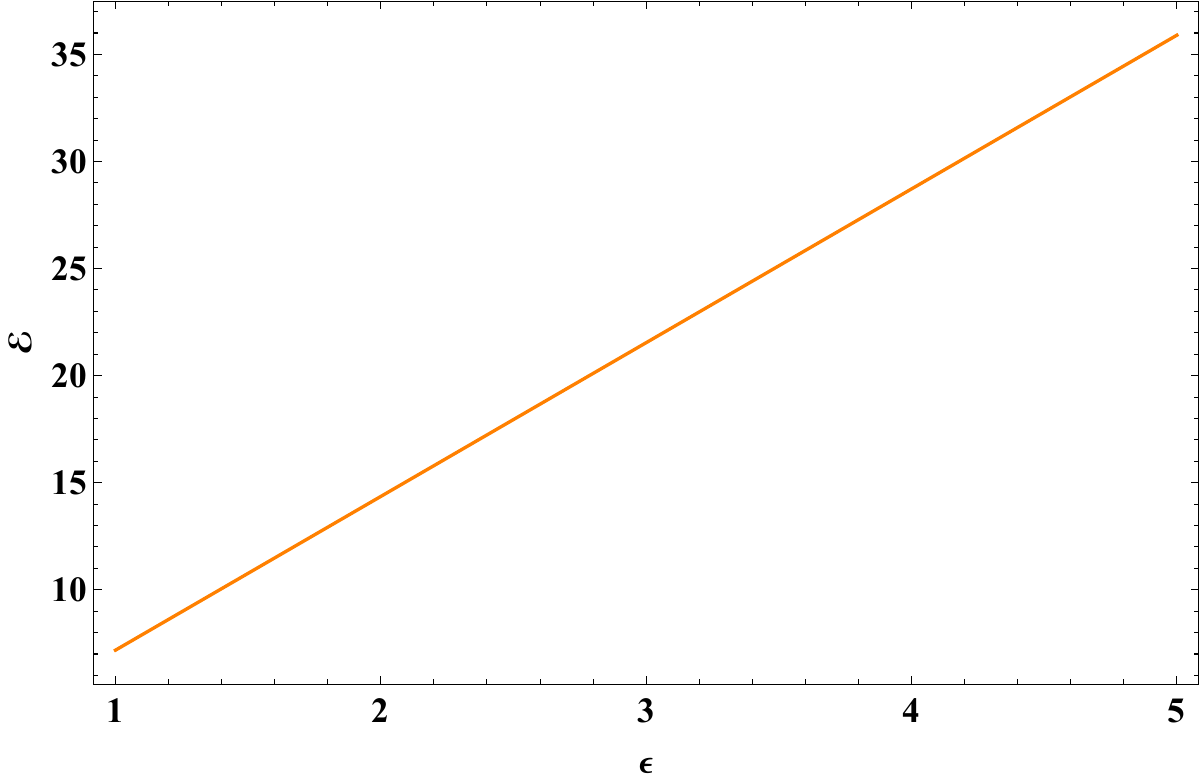}}\,\,\,\,\,\,\,\,\,\,\,\,\,\,\,\,\,\,\,\,\,
       \subfigure[]{\includegraphics[width=6.5cm,height=5cm]{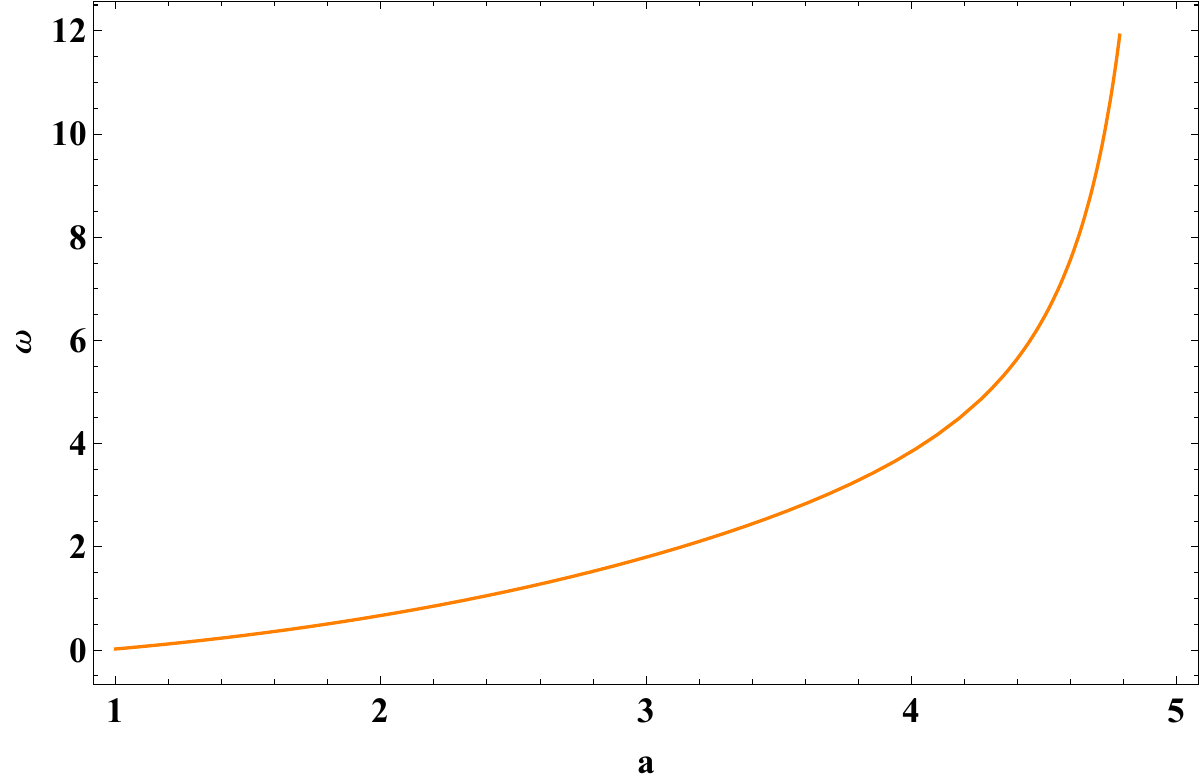}}\,\,\,\,\,\,\,\,\,\,\,\,\,\,\,\,\,\,\,\,\,
        \caption{The figure illustrates the profile of, (a) Variation of proper length $l$ with respect to thickness $\epsilon$, (b)  variation of entropy $\mathcal{S}$ with respect to thickness $\epsilon$, (c)  variation of energy $\mathcal{E}$ with respect to thickness $\epsilon$,(d) variation of EoS $\omega$ with respect to  $\textbf{a}$.}
        \label{fig-6}
    \end{figure}
    
\end{widetext}

Fig-\eqref{fig-6}(d) shows the variation of EoS $\omega$ with respect to $\textbf{a}$.

\section{Boundary Condition} \label{sec:IX}
\begin{itemize}
  \item Reissner-Nordstrom metric:  

\begin{equation}
    1+ \frac{b(r^4-16 c_1)}{4 a r^2}= 1-\frac{2M}{r}+ \frac{Q^2}{r^2},
\end{equation}

\begin{equation}
    e^{c_2} \left(1+\frac{16 c_1}{r^4}\right)=1-\frac{2M}{r}+ \frac{Q^2}{r^2}.
\end{equation}

Obtained constant,
\begin{equation}\label{eq:85}
    c_1= \frac{8 a M r-4 a Q^2+b r^4}{16 b},
\end{equation}
\begin{equation}
    c_2=\log \left(-\frac{b r^2 \left(-2 M r+Q^2+r^2\right)}{2 \left(-4 a M r+2 a Q^2-b r^4\right)}\right),
\end{equation}

\item ABG metric:

\begin{equation}
    1+ \frac{b(r^4-16 c_1)}{4 a r^2}=1-\frac{2M}{r}+ \frac{2M}{r} \tanh{\left(\frac{Q^2}{2 M r}\right)},
\end{equation}
\begin{equation}
     e^{c_2} \left(1+\frac{16 c_1}{r^4}\right)=1-\frac{2M}{r}+ \frac{2M}{r} \tanh{\left(\frac{Q^2}{2 M r}\right)},
\end{equation}

Obtained constant,
\begin{equation}
    c_1=\frac{r \left(-8 a M \tanh \left(\frac{Q^2}{2 M r}\right)+8 a M+b r^3\right)}{16 b},
\end{equation}

\begin{equation}
    c_2=\log \left(\frac{-2 M \tanh \left(\frac{Q^2}{2 M r}\right)+2 M-r}{2 \left(\frac{4 a M \tanh \left(\frac{Q^2}{2 M r}\right)}{b r^2}-\frac{4 a M}{b r^2}-r\right)}\right).
\end{equation}

\item Charged Bardeen metric:

\begin{equation}
    1+ \frac{b(r^4-16 c_1)}{4 a r^2}=1-\frac{2Mr^2}{(r^2+Q^2)^{\frac{3}{2}}}+\frac{Q^2 r^2}{(r+2+Q^2)^2},
\end{equation}
\begin{equation}
     e^{c_2} \left(1+\frac{16 c_1}{r^4}\right)= 1-\frac{2Mr^2}{(r^2+Q^2)^{\frac{3}{2}}}+\frac{Q^2 r^2}{(r+2+Q^2)^2} .
\end{equation}

Obtained constant,
\begin{equation}
    c_1=-\frac{a r^2 \left(-\frac{b r^2}{4 a}-\frac{2 M r^2}{\left(Q^2+r^2\right)^{3/2}}+\frac{Q^2 r^2}{\left(Q^2+r+2\right)^2}\right)}{4 b},
\end{equation}

\begin{equation}
   c_2=\log \left(\frac{-\frac{2 M r^2}{\left(Q^2+r^2\right)^{3/2}}+\frac{Q^2 r^2}{\left(Q^2+r+2\right)^2}+1}{\frac{8 a M}{b \left(Q^2+r^2\right)^{3/2}}-\frac{4 a Q^2}{b \left(Q^2+r+2\right)^2}+2}\right).
\end{equation}

\end{itemize}

\section{Deflection angle for thin shell around gravastar}\label{sec:X}
It is well known that when light passes through a huge body, it behaves like a convex lens. The null geodesic equation changes when light passes through massive masses and along null geodesics due to nontrivial metrics.\\
In 1919, gravitational lensing happened, which was the first experimental test of general relativity. In galaxy cluster studies, weak lensing is turning into a vital instrument.\\
With the absence of an event horizon, this work aims to show that such red-shifted photons can be detected with a next-generation Event Horizon-like radio telescope, provided that the high-energy photon created in the interior region can physically pass through the thin shell. Similar to black holes, we anticipate that the inner region of the gravastar's shadow will be found at some point. The lensing in the context of black holes has been done by Virbhadra and Ellis \cite{Virbhadra/2000} and Bozza \cite{Bozza/2010}. Also, with the advances of the ETH telescope how one can probe the gravitational lensing in the strong field limit \cite{Molla/2024}, which was developed in \cite{Bozza/2002}. We hope that our numerical analysis could be further tasted by the next generation radio telescopes like ETH telescopes.
Multiple methods exist to calculate the deflection angle for any given measure. A particle with mass $m$ and charge $q$ is subject to one of them, which is the application of the Hamilton-Jacobi equation given by\\
$$g^{ij}\left(\frac{\partial S}{ \partial x^i} + q A_i \right) \left(\frac{\partial S}{ \partial x^k} +q A_k \right)+m^2=0$$
We notice that the equation may be integrated for $m=0$ and $q=0$, as we are interested in the photon path. However, if the conventional null geodesic equation is used to get the following metric,
\begin{equation} \label{eq:95}
 ds^2= C(r) dt^2 -D(r) dr^2 -F(r) (d\theta^2+ \sin^2 \theta d\phi^2).  
\end{equation}
By following the instructions given by Virbhadra et al., we calculate the deflection angle as follows. \cite{Virbhadra/1998}.
\begin{equation} \label{eq:96}
    \alpha(r_0)=I(r_0)-\pi,
\end{equation}

where

\begin{equation} \label{eq:97}
   I(r_0)=  \int^{\infty}_{r_0} \frac{2 \sqrt{D(r)}dr}{\sqrt{F(r)}\sqrt{\frac{F(r) C(r_0)}{F(r_0)C(r)}-1}}. 
\end{equation}

\subsection{Reissner-Nordstrom metric:}

For the shell,
\begin{equation}
\begin{gathered}
C(r)=e^{\nu(r)}=\frac{-2 M r+Q^2+r^2}{r^2}, \\
D(r)=e^{\lambda(r)}=\left( \frac{b \left(r^4-\frac{8 a M r-4 a Q^2+b r^4}{b}\right)}{4 a r^2}+1 \right)^{-1}
,\\ F(r)= r^2.
\end{gathered}
\end{equation}

Now, the above equation becomes
\begin{equation}
    I(r_0)_{shell}= 2 \int ^{\infty}_{r_0} \frac{\sqrt{\left( \frac{b \left(r^4-\frac{8 a M r-4 a Q^2+b r^4}{b}\right)}{4 a r^2}+1 \right)^{-1}}}{r \sqrt{\frac{r^2\left(\frac{-2 M r_0+Q^2+r_0^2}{r_0^2}\right)}{r_0^2 \left(\frac{-2 M r+Q^2+r^2}{r^2}\right)}-1}} dr .
\end{equation}

For the exterior region
\begin{multline}
    \begin{gathered}
        C(r)=1-\frac{2M}{r}+ \frac{Q^2}{r^2},\,\,\, D(r)=\left(1-\frac{2M}{r}+ \frac{Q^2}{r^2}\right)^{-1},\\
        F(r)=r^2.
    \end{gathered}
\end{multline}

\begin{equation}
  I(r_0)_{outside} =  2 \int ^{\infty}_{r_0} \frac{\sqrt{\left(1-\frac{2M}{r}+ \frac{Q^2}{r^2}\right)^{-1}}}{r \sqrt{\frac{r^2 \left(1-\frac{2M}{r_0}+\frac{Q^2}{r_0^2}\right)}{r_0^2 \left(1-\frac{2M}{r}+\frac{Q^2}{r^2}\right)}-1}}dr.
\end{equation}
\subsection{ABG metric}

For the shell,
\begin{equation}
    \begin{gathered}
        C(r)=\frac{2 M \tanh \left(\frac{Q^2}{2 M r}\right)-2 M+r}{r},\\ D(r)=\left(\frac{2 M \tanh \left(\frac{Q^2}{2 M r}\right)-2 M+r}{r}\right)^{-1},\\ F(r)=r^2.
    \end{gathered}
\end{equation}

\begin{equation}
    I(r_0)_{shell}= 2\int ^{\infty}_{r_0} \frac{\sqrt{\left(\frac{2 M \tanh \left(\frac{Q^2}{2 M r}\right)-2 M+r}{r}\right)^{-1}}}{r \sqrt{\frac{r^2\left(\frac{2 M \tanh \left(\frac{Q^2}{2 M r_0}\right)-2 M+r_0}{r_0}\right)}{r_0^2 \left(\frac{2 M \tanh \left(\frac{Q^2}{2 M r}\right)-2 M+r}{r}\right)}-1}}  dr.
\end{equation}

For the exterior region, 
\begin{equation}
    \begin{gathered}
        C(r)=1-\frac{2M}{r}+ \frac{2M}{r} \tanh{\left(\frac{Q^2}{2Mr}\right)},\\
        D(r)=\left(1-\frac{2M}{r}+ \frac{2M}{r} \tanh{\left(\frac{Q^2}{2Mr}\right)}\right)^{-1}, F(r)=r^2.
    \end{gathered}
\end{equation}

\begin{equation}
    I(r_0)_{outside}= 2\int ^{\infty}_{r_0} \frac{\sqrt{\left(1-\frac{2M}{r}+ \frac{2M}{r} \tanh{\left(\frac{Q^2}{2Mr}\right)}\right)^{-1}}}{r \sqrt{\frac{r^2 \left(1-\frac{2M}{r_0}+ \frac{2M}{r_0} \tanh{\left(\frac{Q^2}{2Mr_0}\right)}\right)}{r_0^2 \left(1-\frac{2M}{r}+ \frac{2M}{r} \tanh{\left(\frac{Q^2}{2Mr}\right)}\right)}-1}} dr.
\end{equation}

\subsection{Charged Bardeen metric}

For the shell
\begin{equation}
\begin{gathered}
C(r)=\frac{Q^4+Q^2 (r (r+2)+4)+(r+2)^2}{\left(Q^2+r+2\right)^2}-\frac{2 M r^2}{\left(Q^2+r^2\right)^{3/2}},\\
D(r)=\left(r^2 \left(\frac{Q^2}{\left(Q^2+r+2\right)^2}-\frac{2 M}{\left(Q^2+r^2\right)^{3/2}}\right)+1\right)^{-1},\\
F(r)=r^2.
\end{gathered}
\end{equation}

 
\begin{multline}
     I(r_0)_{shell}= 2 \int ^{\infty}_{r_{0}} r\left( \frac{r^2 \mathcal{A}}{r_0^2 \mathcal{B}}-1\right)^{-\frac{1}{2}}\\
     \times \left(r^2 \left(\frac{Q^2}{\left(Q^2+r+2\right)^2}-\frac{2 M}{\left(Q^2+r^2\right)^{3/2}}\right)+1\right)^{-\frac{1}{2}}  dr
 \end{multline}
 
Where $\mathcal{A}=\frac{Q^4+Q^2 (r_0 (r_0+2)+4)+(r_0+2)^2}{\left(Q^2+r_0+2\right)^2}-\frac{2 M r_0^2}{\left(Q^2+r_0^2\right)^{3/2}}$ and $\mathcal{B}=\frac{Q^4+Q^2 (r (r+2)+4)+(r+2)^2}{\left(Q^2+r+2\right)^2}-\frac{2 M r^2}{\left(Q^2+r^2\right)^{3/2}}$

\begin{widetext}

    \begin{figure}[h]
        \centering
        \subfigure[]{\includegraphics[width=5.1cm,height=4.5cm]{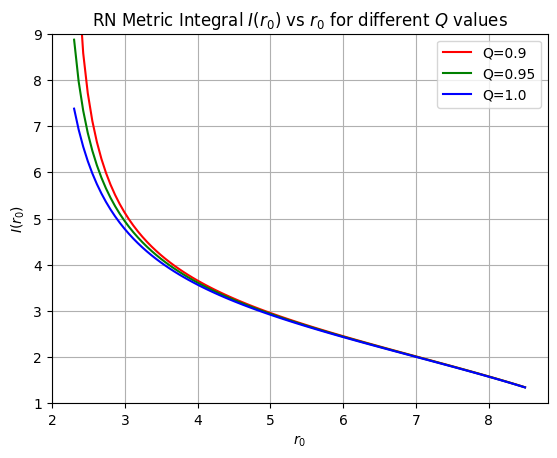}}\,\,\,\,\,\,\,\,\,\,\,\,\,\,\,\,\,\,\,\,\,\,\,\,\,\,
        \subfigure[]{\includegraphics[width=5.1cm,height=4.5cm]{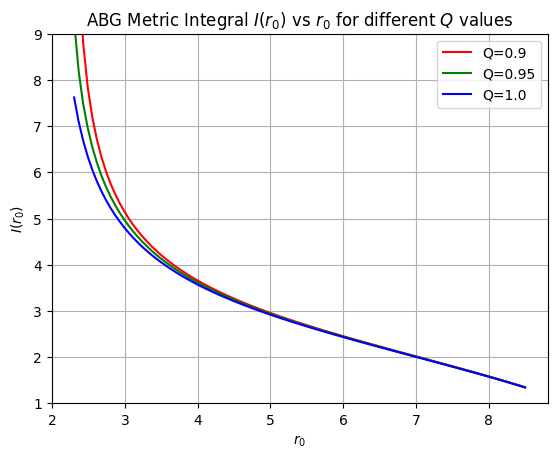}}\,\,\,\,\,\,\,\,\,\,\,\,\,\,\,\,\,\,\,\,\,
        \subfigure[]{\includegraphics[width=5.1cm,height=4.5cm]{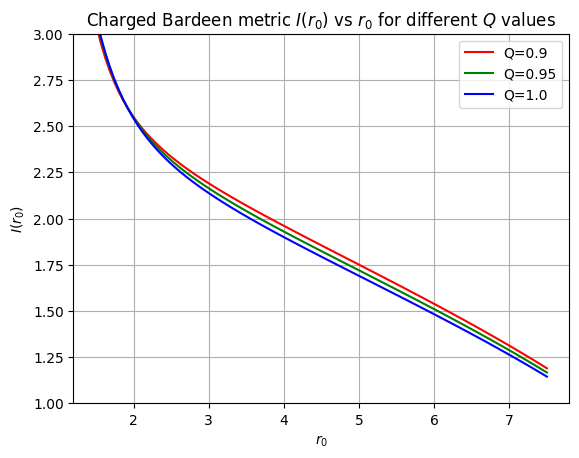}}
        \caption{ The figure illustrates the profile of light deflection angle $I(r_0)$ relative to the radial coordinate $r_0$ for various values of $Q$ on (a)  R-N metric, (b)  ABG metric, and (c)  charged Bardeen metric. }
        \label{fig-7}
    \end{figure}
    
\end{widetext}

For the exterior region,
\begin{multline}
 C(r)=1-\frac{2Mr^2}{(r^2+Q^2)^{\frac{3}{2}}}+\frac{Q^2 r^2}{(r+2+Q^2)^2},
 \\
     D(r)=\left(1-\frac{2Mr^2}{(r^2+Q^2)^{\frac{3}{2}}}+\frac{Q^2 r^2}{(r+2+Q^2)^2}\right)^{-1},
     \\
     F(r)=r^2.   
\end{multline}
\begin{equation}
    I(r_0)_{outside}=2 \int ^{\infty}_{r_0} \frac{\sqrt{\left(1-\frac{2Mr^2}{(r^2+Q^2)^{\frac{3}{2}}}+\frac{Q^2 r^2}{(r+2+Q^2)^2}\right)^{-1}}}{r\sqrt{\frac{r^2\left(1-\frac{2Mr_0^2}{(r_0^2+Q^2)^{\frac{3}{2}}}+\frac{Q^2 r_0^2}{(r_0+2+Q^2)^2}\right)}{r_0^2\left(1-\frac{2Mr^2}{(r^2+Q^2)^{\frac{3}{2}}}+\frac{Q^2 r^2}{(r+2+Q^2)^2}\right)}-1}} dr.
\end{equation}
Therefore, it's clear that $I(r_0)_{outside}$ resembles a typical black hole in its formulation. Yet, the interior varies as it lacks an event horizon. If a red-shifted photon were able to be bent through the shell from the inside, it leads us to expect that forthcoming radio telescopes could differentiate between a black hole and a gravastar.\\
As shown in the Fig \eqref{fig-7} we have done the numerical integration of the $I(r_0)$ for various values of the $Q$ (as shown in the plot). We can see that the results are very similar to the works that has been done earlier in the context of black holes \cite{Manna/2018} and wormholes \cite{Godani/2021}.

\section{Stability} \label{sec:XI}
\begin{widetext}

    \begin{figure}[h]
        \centering
       \subfigure[]{\includegraphics[width=5.1cm,height=4.5cm]{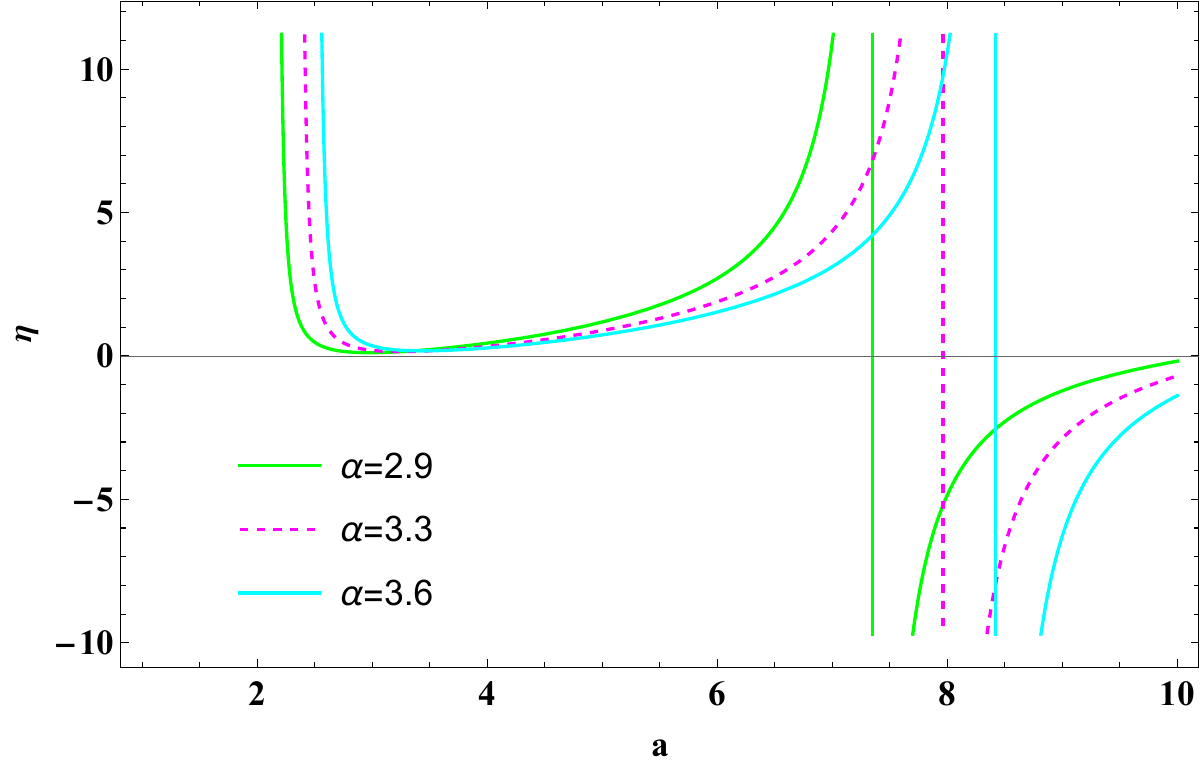}}\,\,\,\,\,\,\,\,\,\,\,\,\,\,\,\,\,\,\,\,\,\,\,\,\,\,
       \subfigure[]{\includegraphics[width=5.1cm,height=4.5cm]{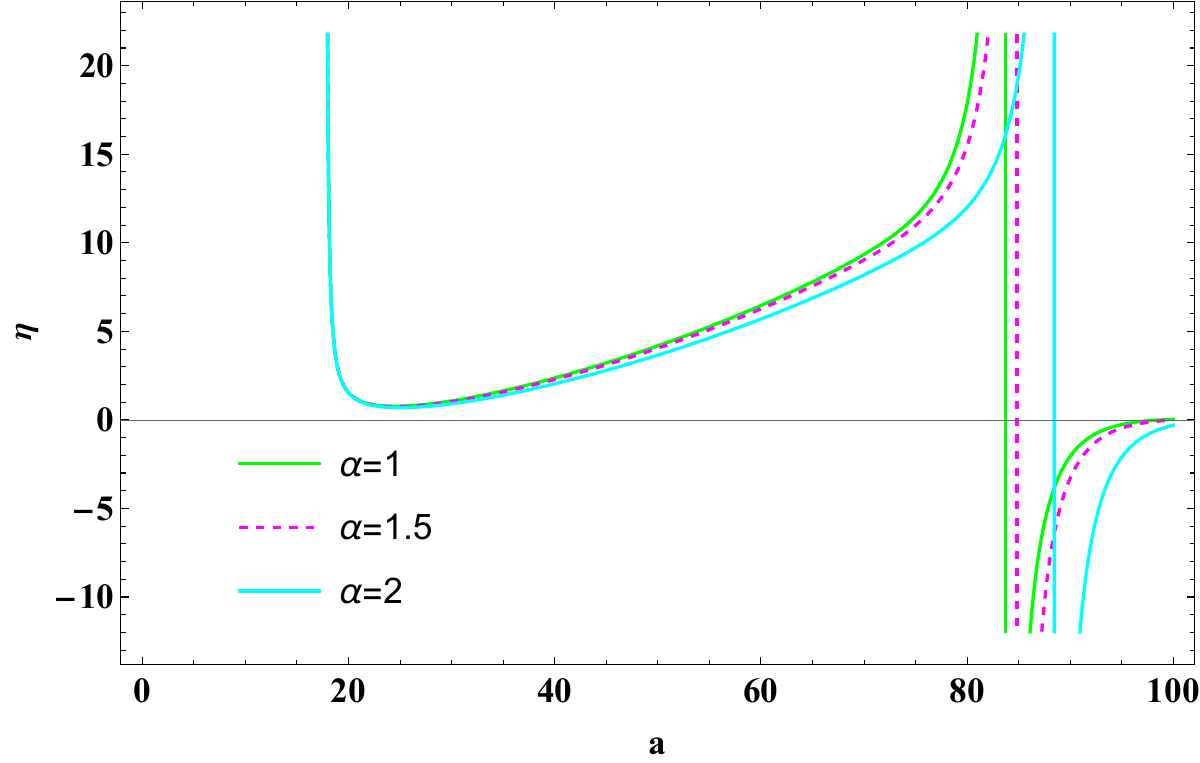}}\,\,\,\,\,\,\,\,\,\,\,\,\,\,\,\,\,\,\,\,\,
       \subfigure[]{\includegraphics[width=5.1cm,height=4.5cm]{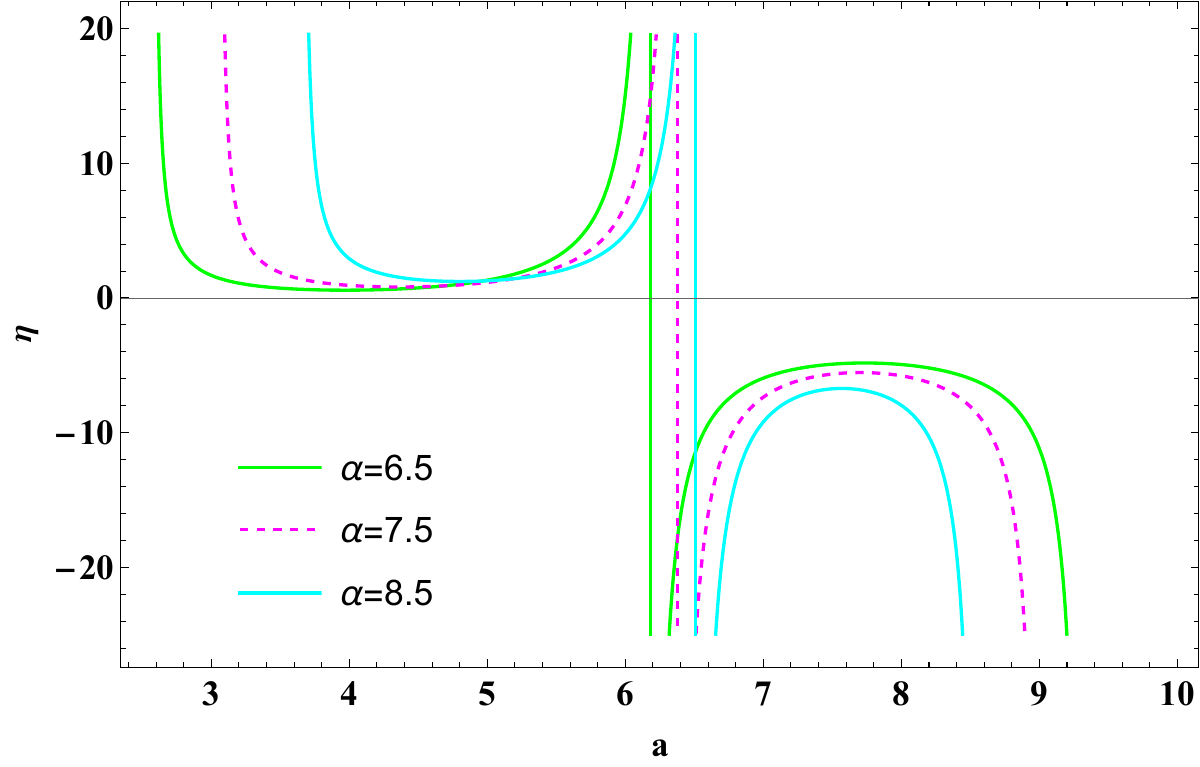}}
        \caption{Variation of $\eta$ with respect $a$ for different values of  $\alpha$ in R-N metric, (b)  ABG metric, (c)  charged Bardeen metric. }
        \label{fig-8}
    \end{figure}
    
\end{widetext}

In the previous sections, we have commented on how we can use the Israel junction conditions to find the physics properties like pressure \textbf{($p$)} energy density\textbf{ $(\zeta)$ } and potentials due to the thin-shell. In Section \ref{sec:IX}, we also discussed how we can utilize this potential to determine the deflection angle of light. This, in turn, can be employed as a phenomenological approach to studying the observation of gravastars. \\
In this section, we dive into another phenomenological study of the thin shell, that is, the stability of the thin shell. Even though the first study of thin shell stability around the black hole was done by Poisson and Visser \cite{Poisson/1995}. The core idea was once we get the potential of the shin shell from Israel junction conditions, one can use the ``shape" of the potential to determine the stability of the thin shell. In this case the minima of potential at $a_0$, that is $V^{\prime}(a_0)=0$ and $V^{\prime\prime}(a_0)\geq 0$, would give the stability conditions.\\
However, there are some inherent problems with such an approach. First of all, it is not thermodynamically well motivated, and the thin potential does not give any information about the matter it is made of, so this definition is not useful for universal purposes.\\
The definition of stability, which indeed captures the thermodynamics essence of the thin shell, was done in the same paper by Poisson and Visser \cite{Poisson/1995} and was further extended by Lobo \cite{Lobo/2003}  and also applied in various contexts in \cite{Pramit/2021, Ovgun/2017,Yousaf/2019}.\\
In this approach, one takes the square of the effective speed of sound ($\eta$) as $\eta=\frac{p^{\prime}}{\varsigma^{\prime}}$ and note that in order to maintain the causality, one needs $0\leq\eta\leq1$, so this gives the stability condition for the thin shell of the gravastar.\\
In figure \eqref{fig-8}, we show the general properties of the $\eta$ with respect to the thickness; note that the singularities appear to show that the formula is not valid at those points (either the adiabatic conditions are violated or the energy density is constant). In figure \eqref{fig-8}, we have shown the stable region of thickness for which $0\leq\eta\leq1$ is satisfied for various values of the model parameter ($\alpha$).\\
However, we would like to note that such a study of stability is not without any limitations. As discussed in Poisson and Visser \cite{Poisson/1995}, the formula for the speed of sound is only valid as long as we can treat it as perfect fluid for some reason. If those conditions happen to break down, we do not have a full understanding of the speed of sound in those contexts. So, we can only say the above regions show sufficient conditions for stability, which is not necessary.

\begin{widetext}

    \begin{figure}[H]
        \centering
       \subfigure[]{\includegraphics[width=5cm,height=4.5cm]{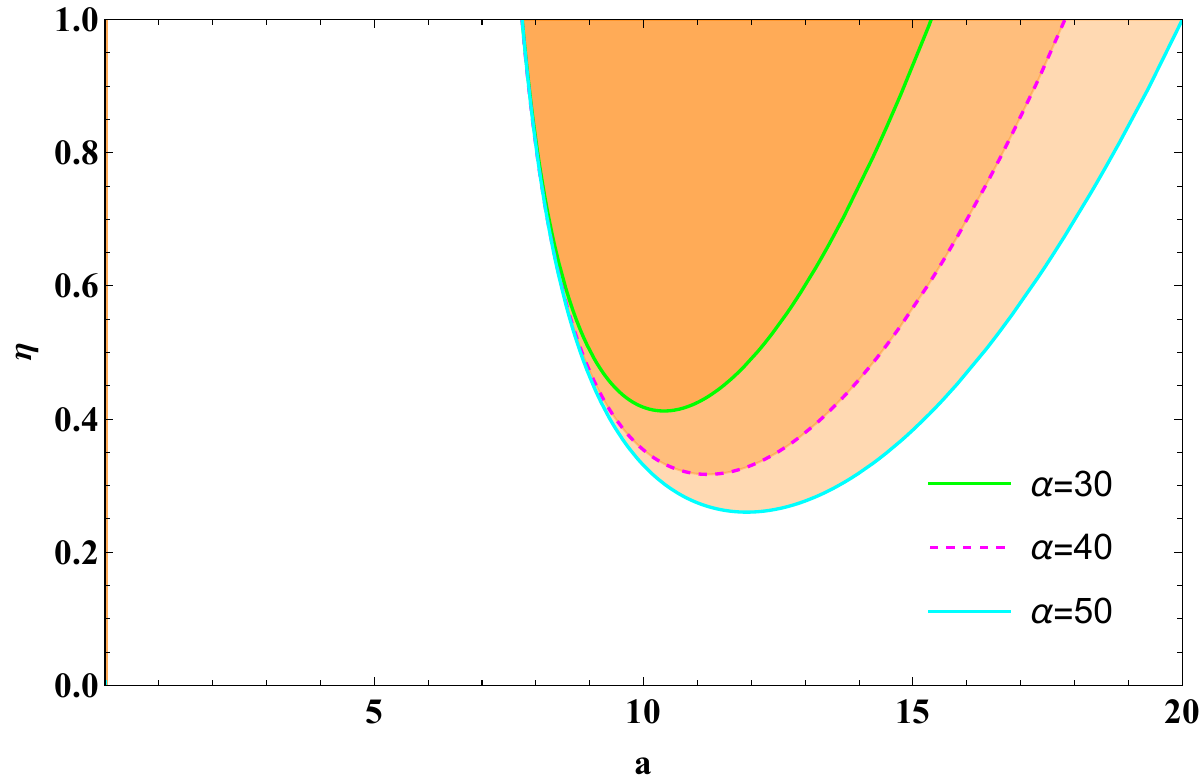}}\,\,\,\,\,\,\,\,\,\,\,\,\,\,\,\,\,\,\,\,\,\,\,\,\,\,
       \subfigure[]{\includegraphics[width=5cm,height=4.5cm]{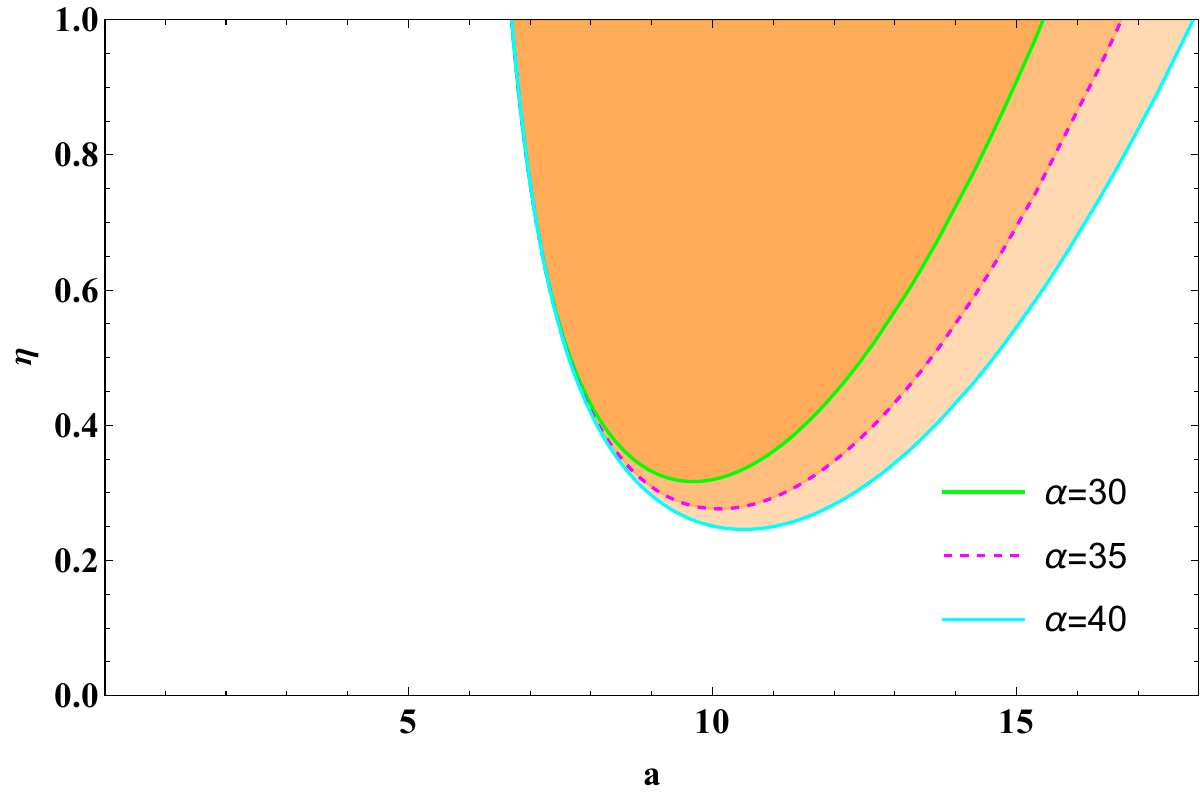}}\,\,\,\,\,\,\,\,\,\,\,\,\,\,\,\,\,\,\,\,\,
       \subfigure[]{\includegraphics[width=5cm,height=4.5cm]{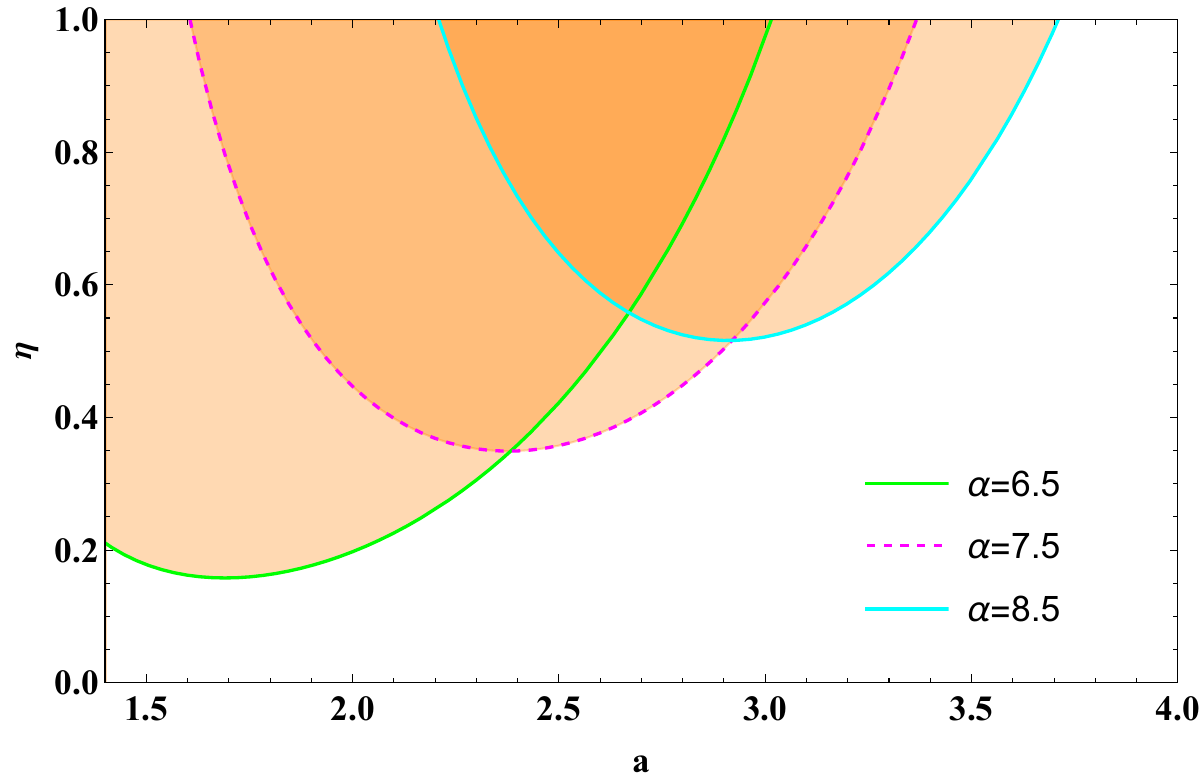}}
        \caption{Variation of $\eta$ with respect $a$ for different values of  $\alpha$ in R-N metric, (b)  ABG metric, (c)  charged Bardeen metric. }
        \label{fig-9}
    \end{figure}
    
\end{widetext}

\section{Conclusions}\label{sec:XII}
 In this article, we have studied the charged gravastar in noncommutative geometry in $f(\mathbb{T})$ gravity. In section \ref{sec:I}, we have given a brief introduction to the effect of noncommutative geometry on energy density. We have briefly demonstrated how the path-integral approach can lead to such an energy density given in equation \eqref{eq:1}. We have also given a brief overview of the current and past work on various astrophysical objects that have been explored in the context of $f(\mathbb{T})$ gravity and noncommutative geometry. \\
 In section \ref{sec:II}, we use the Einstein-Hilbert action in the form of $\mathbb{T}$ and note that under which condition it reduces to GR (i.e. $f(\mathbb{T})=\mathbb{T}$), we write the field equation from the variation principle and also define some of the relevant terms like torsion ($\mathbb{T}^{\sigma} \,_{\mu \nu}$),
contorsion ($ K^{\mu \nu}\,_{\sigma}$)   etc. along the way.\\
In section \ref{sec:III}, we briefly recap the derivation of the field equation for static spherically symmetric metric under $f(\mathbb{T})$ gravity. We derive all the necessary terms like torsion ($\mathbb{T}^{\sigma} \,_{\mu \nu}$) ,
contorsion ($ K^{\mu \nu}\,_{\sigma}$) and calculate the full field equation \eqref{eq:7} using the diagonal tetrads $e^i\,_{\mu}$. We also like to note that due to equation\eqref{eq:18}, we have strong constraints on the form of $f(\mathbb{T})$ we are allowed to choose. As it is evident from equation \eqref{eq:18}, we can only take the linear model, and we have taken $f(\mathbb{T})=a\mathbb{T}+b$. Finally, we write the field equation for our metric in equation \eqref{eq:20}-\eqref{eq:22} and the conservation equation in \eqref{eq:24}.\\
In section \ref{sec:IV}, we have focused on the interior of the gravastar where the EoS $(\omega)=-1$, and we have also incorporated the noncommutative geometry only in the interior of the gravastar. We use the metric of the charged noncommutative black hole solution in the interior to find the metric potentials. We have also used $\rho=-p$ to get the full solution in the interior.\\
In section \ref{sec:V}, we have concentrated our attention on the shell of the gravastar. As we have discussed earlier in the introduction how having stiff matter ($\omega=1$) in the shell makes the gravastar entropy coincide with the black hole entropy. In this section, we have solved for shell taking some appropriate limit of the thin shell ($0<e^{-\lambda}<<1$).\\
In section  \ref{sec:VI}, we focused our attention on the exterior of the charged black holes. We have taken three different types of metrics outside that: A. R-N metric, B. ABG metric, and C. Charged Bardeen metric. We have briefly discussed the theoretical and phenomenological motivation for all three metrics.\\
In section \ref{sec:VII}, we have dived into the formulation of Israel junction conditions on the shell. We have calculated the energy density ($\varsigma$) and pressure ($p$) for the thin shell using the junction condition, which we later use for the stability analysis of the thin shell. We have also shown how one can use the junction condition to derive the potential $V(\textbf{a})$ in equation \eqref{eq:56.1} for the thin shell. Fig \eqref{fig-5} shows the profile of the potential. We can note two things: the potential has definite minima, which means it is stable following the discussion given by Poisson and Visser \cite{Poisson/1995}. We would also like to note that such minima of potential can be used to find the innermost stable circular orbit (ISCO) of the shell, which in principle can be used to check the properties of the accretion disk near the gravastar, and one can observationally distinguish both. So, this potential profile has some serious phenomenological values. \\
 In section \ref{sec:VIII}, we derive various physical properties which are relevant to the phenomenological properties of the gravastars. For example, In subsection A, proper length $(l)$, which we calculate by noting in the thin shell approximation, can be used in the Taylor series expansion to find the integral (which does not have an exact analytical solution). Fig \eqref{fig-6}- (a) shows the variation of the proper length ($l$) with respect to the thickness ($\epsilon$). In subsection B, we calculate the entropy ($\mathcal{S}$) of the gravastar for this system. We note that as in the interior $\rho = -p$, the only contribution for the entropy that is coming is from the shell. We have again used the Taylor series expansion for calculating the integral and plotted the entropy ($\mathcal{S}$) with respect to the thickness ($\epsilon$) in  Fig \eqref{fig-6}- (b). In subsection C, we have calculated the energy ($E$) of the shell from the energy density ($\rho$) we have calculated using the Israel junction condition.  Fig \eqref{fig-6}- (c) shows the plot of energy ($E$) vs thickness ($\epsilon$). Finally, in subsection D, we have calculated the effective EoS ($\omega=\frac{p}{\varsigma}$) for the thin shell and plotted with respect to the $a$. We note that it is going beyond the stiff matter limit, so we studied the stability analysis in the later section to show the allowed values of $a$. \\
 In section \ref{sec:IX}, we explicitly calculate the integration constants like $c_1$, $c_2$, etc., from the boundary condition, which we use in the later sections.\\
In Section \ref{sec:X}, we present a phenomenological analysis of the gravastar by calculating the deflection angle ($\alpha(r_0)$) induced by the surrounding shell. We anticipate that future advancements in radio telescopes, such as those developed by EHT, will be capable of detecting the distinctive signatures of gravastars, thus enabling differentiation from black holes. In Fig \eqref{fig-7} we have plotted the deflection angle $(I(r_0))$ vs $(r_0)$ for different values of charge, which will much more suitable for current research and that this could indeed be checked by ETH or next generation radio telescopes in the future.\\
In section \ref{sec:XI}, we do the stability analysis of the gravastar. We first start with a brief introduction to why stability analysis is necessary, followed by a brief literature survey. After that, we do the thermodynamical stability analysis by using the speed of sound square ($\eta$) is given by $\eta=\frac{p^{\prime}}{\varsigma^{\prime}}$, in Fig \eqref{fig-8} we have plotted square of the speed of sound ($\eta$) vs the shell radius ($a$) (for all three metrics with different $\alpha$), also has given physical motivation for the singularities. Finally, Fig \eqref{fig-9} gives the allowed values of $a$ for which the stability criteria ($0<\eta<1$) is satisfied. We have also given the limitations of such stability analysis and how these can be improved in the future. \\
In future, we hope that this analysis can be extended into more general $f(T,\mathcal{T})$ gravity \cite{Ghosh/2020},$f(Q, T)$ gravity \cite{Pradhan/2023},$f(\mathcal{G})$ \cite{Bhatti/2021} gravity, $f(R, T)$  \cite{Das/2017} classes of gravity as well. Also, in the near future, with more EHT observation data, we will be able to distinguish gravastar from black holes from the gravitational lensing.

\section*{Data Availability}

There are no new data associated with this article.

\acknowledgements SG acknowledges Council of Scientific and Industrial Research (CSIR), Government of India, New Delhi, for junior research fellowship (File no.09/1026(13105)/2022-EMR-I). PKS  acknowledges the National Board for Higher Mathematics (NBHM) under the Department of Atomic Energy (DAE), Govt. of India for financial support to carry out the Research project No.: 02011/3/2022 NBHM(R.P.)/R \& D II/2152 Dt.14.02.2022 and IUCAA, Pune, India for providing support through the visiting Associateship program.


\begin{thebibliography}{50}%
\bibitem{Oppenheimer/1939} J.  R.  Oppenheimer,  H. Snyder, \textit{ Phys.Rev.D}  \textbf{56}, 455 (1939).
\bibitem{Penrose/1965} R. Penrose, \textit{Phys. Rev. Lett.} \textbf{14}, 57 (1965).

\bibitem{Akiyama/2019} K.  Akiyama et al., \textit{ Astrophys. J. Lett.} \textbf{875}, L5 (2019).

\bibitem{Landea/2016} I. S. Landea et al., \textit{Phys. Rev. D} \textbf{94}, 104006 (2016).

\bibitem{Alcubierre/2023} M. Alcubierre et al., \textit{Phys. Rev. D} \textbf{107}, 045017 (2023).

\bibitem{Kumar/2016} S.  Kumar et al., \textit{Phys. Rev. D}  \textbf{93}, 101501 (2016).


\bibitem{P/2023} P. O.  Mazur, E.  Mottola, \textit{Universe} \textbf{9}, 88 (2023).

\bibitem{Mazur/2004} P.  O.  Mazur, E.  Mottola, \textit{Proc. Natl. Acad. Sci.} \textbf{101}, 9545 (2004).
\bibitem{Nicolini/2006} P.  Nicolini et al., \textit{Phys. Lett. B}  \textbf{632}, 547-551 (2006). 


\bibitem{Smailagic1/2003} A.  Smailagic , E.  Spallucci, \textit{J. Phys. A} \textbf{36}, L467 (2003).
\bibitem{Ansoldi/2007} S.  Ansoldi et al., \textit{ Phys. Lett. B} \textbf{645}, 261-266 (2007).


\bibitem{Rizzo/2006} T.  G.  Rizzo, \textit{J. High Energy Phys.}  \textbf{2006}, 021 (2006).
\bibitem{Ghosh/2018} S.  G.  Ghosh, \textit{Class. Quantum Gravity}  \textbf{35}, 085008 (2018).

\bibitem{Banerjee/2008} R.  Banerjee et al., \textit{Phys. Rev. D} \textbf{77}, 124035 (2008).


\bibitem{Spallucci} A.  Smailagic , E. Spallucci, \textit{J. Phys. A: Math. Gen.} \textbf{37}, 7169 (2004).


\bibitem{Doplicher} S. Doplicher et al., \textit{Phys. Lett. B} \textbf{331}, 39 (1994).

\bibitem{Gangopadhyay/2009} S. Gangopadhyay et al ., \textit{Phys. Rev. Lett.} \textbf{102}, 241602 (2009).


\bibitem{Nicolini} P. Nicolini, \textit{Int. J. Mod. Phys. A} \textbf{24}, 1229-1308 (2009).

\bibitem{Nicolini2} P. Nicolini , E. Spalluci, \textit{Class. Quant. Grav.} \textbf{27}, 015010 (2010).
\bibitem{Islam} F. Rahaman et al., \textit{Phys. Rev. D} \textbf{86}, 106010 (2012).

\bibitem{Banerjee} F. Rahaman et al., \textit{Int. J. Theor. Phys.} \textbf{53}, 1910 (2014).
\bibitem{Hassan1} Z.  Hassan et al.,  \textit{Symmetry} \textbf{13}, 1260 (2021).

\bibitem{Tayade/2023} M. Tayade et al., \textit{Phys.Dark Univ.} \textbf{42},  101288 (2023).

\bibitem{Eloy/1999} E.  Ayon-Beato , A.  Garcla , \textit{Phys. Lett. B} \textbf{464}, 25-29 (2021).


 
\bibitem{Ayon/1998} E.  Ayon-Beato , A.  Garcia, \textit{Phys. Rev. Lett. } \textbf{80}, 5056 (1998).
\bibitem{Bronnikov/2001} K.  A.  Bronnikov, \textit{Phys. Rev. D} \textbf{63},  044005 (2001).




\bibitem{Matyjasek/2013} J. Matyjasek et al., \textit{Phys. Rev. D} \textbf{87}, 124025 (2013).

\bibitem{Nicolini/2018} P. Nicolini, \textit{Phys. Lett. B} \textbf{778}, 88-93 (2018).
\bibitem{Ovgun/2020} A. Ovgun et al., \textit{Mod. Phys. Lett. A} \textbf{35}, 2050163 (2020).

\bibitem{Araujo/2024} A.  A.  Araujo Filho et al., \textit{Phys.Dark Univ.} \textbf{46}, 101630 (2024).

\bibitem{Das/2020} A.  Das et al., \textit{Nucl. Phys. B } \textbf{954},  114986 (2020).

\bibitem{Pradhan/2023} S. Pradhan et al., \textit{Chin. Phys. C }\textbf{47}, 095104 (2023).

\bibitem{Mohanty} D.  Mohanty et al., \textit{Ann. Phys.} \textbf{463}, 169636 (2024).

\bibitem{Ovgun/2017} A. Ovgun et al., \textit{Eur. Phys. J. C}  \textbf{77}, 1-7 (2017).


\bibitem{Lobo/2013} F. S. N.  Lobo, R.  Garattini, \textit{	J. High Energy Phys.} \textbf{12}, 065 (2013).

\bibitem{Debnath/2019} U.  Debnath , \textit{Eur. Phys. J. C} \textbf{79}, 1-9 (2019).

\bibitem{Bengochea/2009} G.  R.  Bengochea, R. Ferraro,\textit{Phys. Rev. D} \textbf{79}, 124019 (2009). 
\bibitem{Baojiu/2011}  B. Li  et al., \textit{Phys. Rev. D}  \textbf{83}, 064035 (2011).

\bibitem{Bohmer/2011} C.  G.  Bohmer et al., \textit{Class. Quantum Gravity } \textbf{28}, 245020 (2011).

\bibitem{Zeldovich/1972} Y.  B.  Zeldovich, \textit{Mon. Not. R. Astron. Soc.} \textbf{160}, 1P-3P (1972).

\bibitem{Madsen/1992} M.  S.  Madsen et al., \textit{Phys. Rev. D} \textbf{46}, 1399 (1992).

\bibitem{Carr/1975} J.  B.  Carr, \textit{Astrophys. J.}\textbf{201}, 1-19 (1975).

\bibitem{Braje/2002} T.  M.  Braje et al., \textit{Astrophys. J.} \textbf{580}, 1043 (2002).

\bibitem{Rahaman/2014} F. Rahaman et al., \textit{Eur. Phys. J. C} \textbf{74}, 2845 (2014).

\bibitem{Mohanty/2024} D.  Mohanty , P.  K. Sahoo, \textit{Fortsch.Phys.} \textbf{72}, 2400082 (2024).
 \bibitem{Israel/1966} W. Israel et al., \textit{Nuo. Cim. B} \textbf{44}, 1-14 (1966).


 \bibitem{Rahman/2021} N.  Rahman et al., \textit{Int. J. Mod. Phys. A} \textbf{36}, 2150085 (2021).
 \bibitem{Bardeen/1968} J. Bardeen, \textit{Proceedings of GR5} (Tiflis,USSR,1968).
 
 
\bibitem{Ramadhan/2023} H.  S.  Ramadhan et al., \textit{Eur. Phys. J. C} \textbf{83}, 465 (2023).

\bibitem{Ayon-Beato/2000} E. Ayon-Beato, A.  Garcia, \textit{Phys.Lett. B}  \textbf{493}, 149-152 (2000).

\bibitem{Sen/1924} N. Sen , \textit{Ann. Phys.} \textbf{378}, 365-396 (1924).



\bibitem{Lanczos/1924} K.  Lanczos, \textit{ Ann. Phys.} \textbf{379}, 518--540 (1924).

\bibitem{Darmois/1927} G.  Darmois et al., \textit{Fascicule XXV Gauthier-Villars, Paris,} (1927).


 \bibitem{Israel/1967} W. Israel, \textit{Nuo. Cim. B} \textbf{48}, 463 (1967).
 

 
 \bibitem{Forghani/2020} S.  D.  Forghani et al., \textit{Phys. Lett. B} \textbf{804}, 135374 (2020).
 \bibitem{MSharif} M. Sharif et al., \textit{Chinese J. Phys.} \textbf{65}, 242-253 (2020).
\bibitem{eric} E. Poisson, \textit{A Relativist's Toolkit} (Cambridge University Press, 2007).
 


\bibitem{Virbhadra/2000} K. S. Virbhadra, G.  F. Ellis, \textit{Phys. Rev. D} \textbf{62}, 084003 (2000).

\bibitem{Bozza/2010} V.  Bozza, \textit{Gen. Relativ. Gravit.} \textbf{42}, 2269-2300 (2010).

\bibitem{Molla/2024} N. U.  Molla et al., \textit{Eur. Phys. J. C} \textbf{84}, 574 (2024).

\bibitem{Bozza/2002} V. Bozza, \textit{Phys. Rev. D} \textbf{66}, 103001 (2002).

\bibitem{Virbhadra/1998} K.  S.  Virbhadra et al., \textit{Astron. Astrophys.} \textbf{337}, 1--8 (1998).

\bibitem{Manna/2018} T. Manna et al., \textit{Gen. Relativ. Gravit.} \textbf{50}, 1-18 (2018).

\bibitem{Godani/2021} N. Godani, G. C.  Samanta, \textit{Ann. Phys. } \textbf{429}, 168460 (2021).

\bibitem{Poisson/1995}E.  Poisson , M.   Visser, \textit{Phys. Rev. D} \textbf{52},7318 (1995).

 \bibitem{Lobo/2003} F. S. N. Lobo , P. Crawford, \textit{Class. Quantum Gravity} \textbf{21}, 391 (2003).
\bibitem{Pramit/2021} P. Bhar , P. Rej, \textit{ Eur. Phys. J. C} \textbf{81}, 1-17 (2021).



\bibitem{Yousaf/2019} Z . Yousaf et al., \textit{Phys.Rev.D} \textbf{100}, 024062 (2019).
\bibitem{Seiberg/1999} N.  Seiberg, E. Witten \textit{J. High Energy Phys.} \textbf{1999}, 032 (1999).

\bibitem{Ghosh/2020} S. Ghosh et al., \textit{Int. J. Mod. Phys. A} \textbf{35}, 2050017 (2020).
 \bibitem{Bhatti/2021} M.  Z.  Bhatti et al., \textit{Chin. J. Phys.} \textbf{73}, 167-178 (2021).


\bibitem{Das/2017} A. Das et al., \textit{Phys. Rev. D} \textbf{95}, 124011 (2017).
\end{thebibliography}
\end{document}